\documentclass[pdflatex,sn-mathphys-num]{sn-jnl}
\usepackage{amsmath,amssymb,amsfonts}%
\usepackage{amsthm}%
\newtheorem{definition}{Definition}[section]
\usepackage{makeidx}  
\usepackage{graphicx}
\usepackage{mymacros}
\usepackage[table,xcdraw]{xcolor}
\usepackage{wrapfig}
\usepackage[makeroom]{cancel}
\usepackage{enumerate}
\usepackage{subcaption}
\usepackage{adjustbox}
\usepackage[ruled,vlined,linesnumbered]{algorithm2e}
\usepackage{amsmath,algpseudocode}

\makeatletter
\renewcommand{\Function}[2]{%
	\csname ALG@cmd@\ALG@L @Function\endcsname{#1}{#2}%
	\def\jayden@currentfunction{#1}%
}
\newcommand{\funclabel}[1]{%
	\@bsphack
	\protected@write\@auxout{}{%
		\string\newlabel{#1}{{\jayden@currentfunction}{\thepage}}%
	}%
	\@esphack
}
\makeatother
\algdef{SE}{Begin}{End}{\textbf{begin}}{\textbf{end}}

\usepackage[section]{placeins}
\usepackage{verbatim}
\usepackage{mathabx}
\usepackage{multirow}
\usepackage{capt-of}
\usepackage{tikz}
\usepackage{nameref}
\newcounter{mylabelcounter}

\makeatletter
\newcommand{\labelText}[2]{%
	#1\refstepcounter{mylabelcounter}%
	\immediate\write\@auxout{%
		\string\newlabel{#2}{{1}{\thepage}{{\unexpanded{#1}}}{mylabelcounter.\number\value{mylabelcounter}}{}}%
	}%
}

\algnewcommand\algorithmicforeach{\textbf{for each}}
\algdef{S}[FOR]{ForEach}[1]{\algorithmicforeach\ #1\algorithmicdo}
\def\univs{\mathbb{U}}
\newcommand{\univ}[1]{\univs_{\mathit{#1}}}

\SetKwInput{KwData}{Input}
\SetKwInput{KwResult}{Output}

\SetCommentSty{mycommfont}

\definecolor{darkred}{rgb}{0.6, 0, 0} 
\definecolor{darkblue}{rgb}{0, 0, 0.6} 

\usepackage{manfnt}
\usepackage{tikz}
\usepackage{stackrel}
\usepackage[bottom]{footmisc}

\usepackage{subcaption}

\usepackage{hyperref}
\usepackage{graphicx}

\newcommand{\orcidID}[1]{%
	\href{https://orcid.org/#1}{%
		\includegraphics[height=1.6ex]{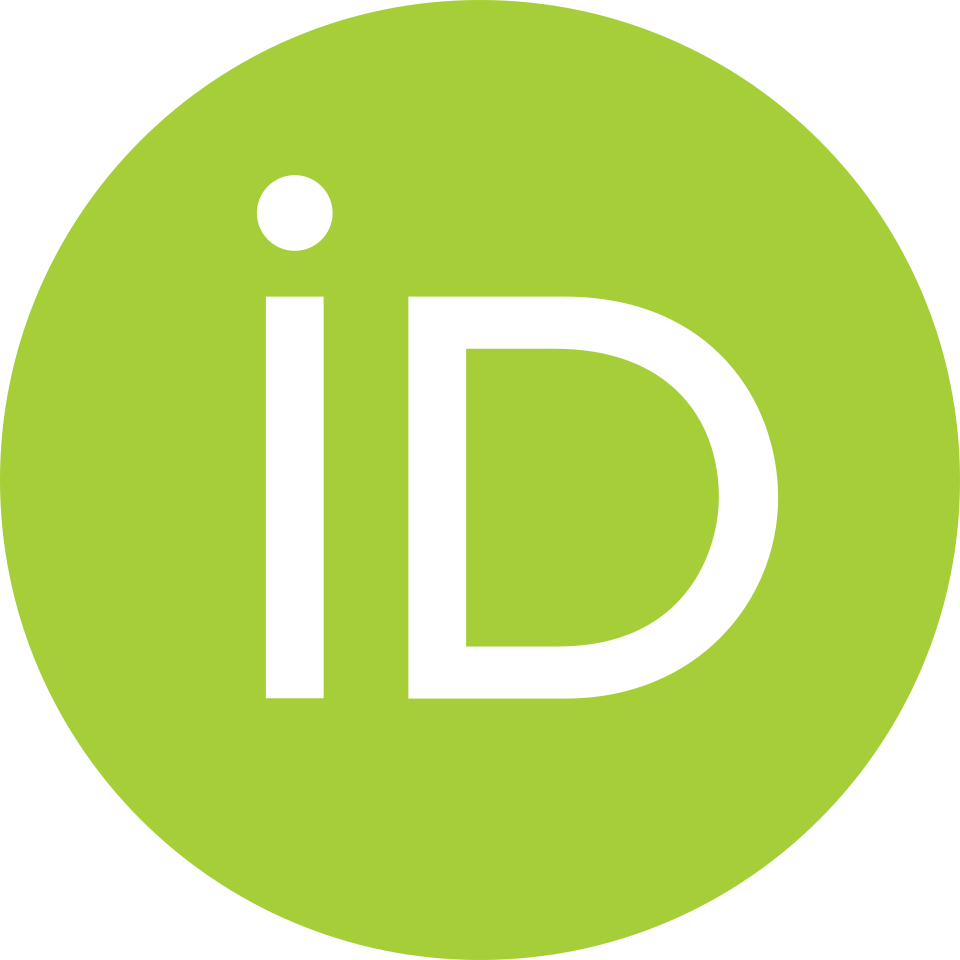}%
	}%
}

\newif\ifreview
\reviewtrue     

\begin{document}

\title[Article Title]{Advancing Object-Centric Process Mining with Multi-Dimensional Data Operations}

\author[1]{\fnm{Shahrzad} \sur{Khayatbashi}\orcidID{0000-0001-7621-0985}}
\email{shahrzad.khayatbashi@liu.se}

\author[2]{\fnm{Najmeh} \sur{Miri}\orcidID{0009-0008-8069-4159}}
\email{najmeh.miri@dsv.su.se}

\author*[2]{\fnm{Amin} \sur{Jalali}\orcidID{0000-0002-6633-8587}}
\email{aj@dsv.su.se}

\affil*[1]{\orgdiv{Department of Computer and Information Science}, \orgname{Link\"oping University}, \orgaddress{\street{M\"aster Mattias}, \city{Link\"oping}, \postcode{581 83}, \state{Link\"oping}, \country{Sweden}}}

\affil[2]{\orgdiv{Department of Computer and System Sciences}, \orgname{Stockholm University}, \orgaddress{\street{Borgarfjordsgatan}, \city{Stockholm}, \postcode{164 25}, \state{Kista}, \country{Sweden}}}

\abstract{Analyzing process data at varying levels of granularity is important to derive actionable insights and support informed decision-making. Object-Centric Event Data (OCED) enhances process mining by capturing interactions among events and multiple objects, leading to the discovery of more detailed and realistic yet complex process models. The lack of methods to adjust the granularity of the analysis prevents users from leveraging the full potential of Object-Centric Process Mining (OCPM). To address this gap, we propose four operations: drill-down, roll-up, unfold, and fold, which enable analysts to change the granularity of analysis when working with Object-Centric Event Logs (OCEL). These operations allow analysts to seamlessly transition between detailed and aggregated process models, facilitating the discovery of insights that require varying levels of abstraction. We formally define these operations and implement them in an open-source Python library. To validate their utility, we applied the approach to real-world OCEL data extracted from a learning management system, covering a four-year period and approximately 400 students, as a case of object-centric educational process mining. This case study shows significant improvements in the precision and fitness of the discovered models after applying the operations. In addition, we evaluate the scalability of the operators on large, publicly available OCELs derived from the Business Process Intelligence Challenge datasets, demonstrating that the operations remain computationally feasible on industrial-scale event logs. This approach can empower analysts to perform more flexible and comprehensive process exploration, unlocking actionable insights through flexible granularity adjustments.}


\maketitle

%
\section{Introduction}\label{Sec:Introduction}
The ability to analyze data at varying levels of granularity is crucial for organizations striving to identify bottlenecks and drive process improvements~\cite{van2021event}. Adapting the level of detail allows users to seamlessly transition between granular views and high-level overviews of business processes. This flexibility enables the discovery of actionable insights that may remain hidden when confined to a single analytical perspective. In complex data environments, dynamic granularity adjustment based on specific analytical goals empowers stakeholders to tailor their analyses, resulting in more precise and effective decision-making.

Object-Centric Event Data (OCED)~\cite{fahland2024oced} offers a richer way to record process data by capturing interactions and dependencies between multiple objects within a single event. This capability surpasses traditional event logs~\cite{van2019object}, which often focus on single-case identifiers. Object-Centric Event Log (OCEL)~\cite{berti2023ocelspecification}, a widely adopted OCED log format~\cite{berti2023pm4py,berti2023oc,adams2022ocpa,liss2024totem,adams2024super,jalali2022object}, associates events with multiple objects. For example, in a hospital setting, the event `register a test' may involve various objects, such as a patient, caregiver, and different diagnostic procedures (e.g., different types of ECG or blood tests). Object-Centric Process Mining (OCPM) enables process analyses from each of these object types' perspectives.

\begin{figure}[t]
	\includegraphics[width=1\textwidth]{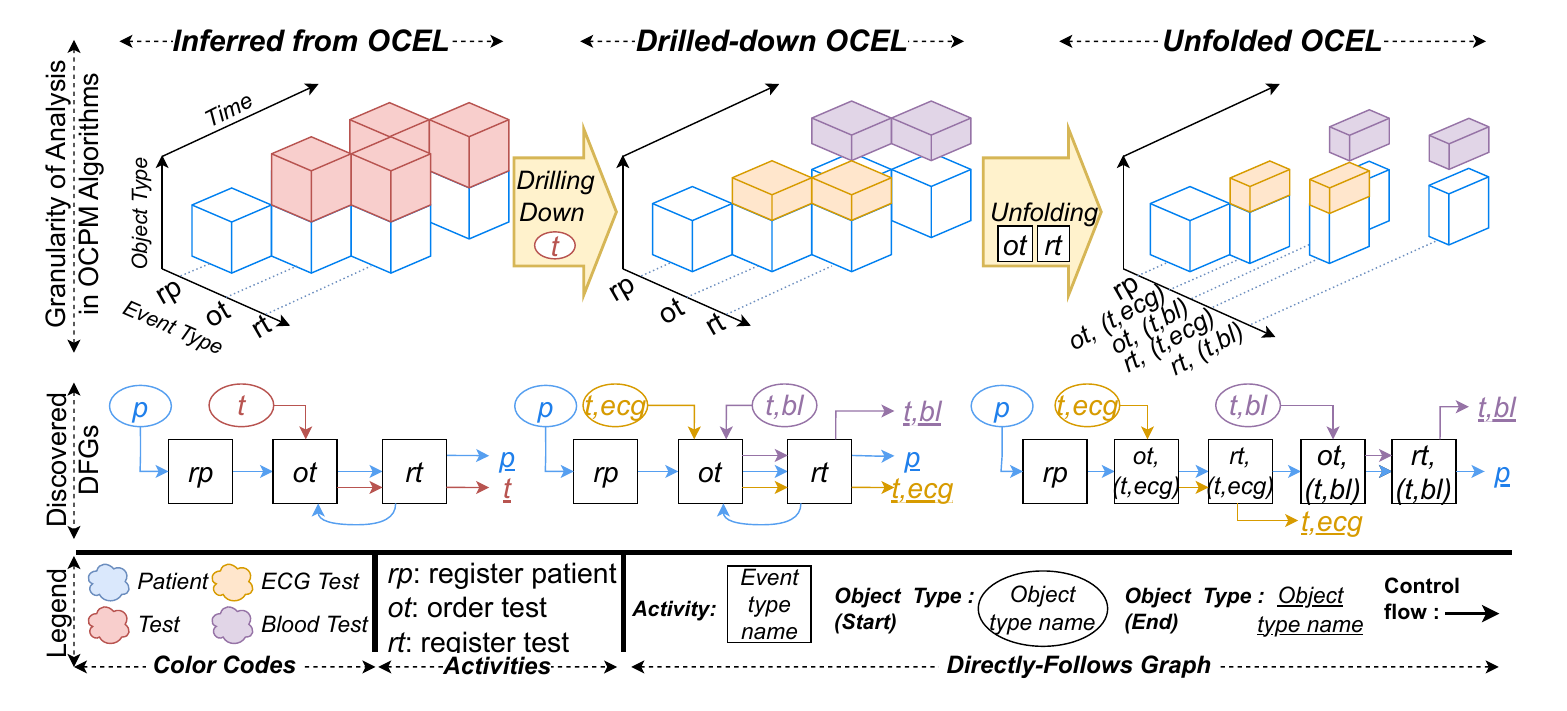}
	\caption{The use of Drill-down and Unfold operations to enable identifying more detailed process patterns in OCPM.}
	\label{fig:runningexample}
\end{figure}

Most OCPM algorithms operate at a higher level of abstraction, deriving process logic for \texttt{Event Types} based on the sequence of events that occurred for each \texttt{Object Type} over \texttt{Time}. This abstraction is illustrated in the upper-left of \figurename\ref{fig:runningexample}, which serves as a running example throughout this paper.
The inferred process logic enables discovering process models; for example, PM4Py can discover an Object-Centric Directly-Follows Graph (OC-DFG)~\cite{berti2023pm4py}, which visualizes directly-follows relationships in the process, as shown in the lower left side of the figure.

Our running example is about the `Chest Pain Evaluation' process in a hospital. The process begins with the registration of a patient (\texttt{rp}), followed by the ordering of an ECG test (\texttt{ot}), which is documented when the results are registered (\texttt{rt}). In practice, ordering different sorts of tests and registering their results produces events of the same type. This behavior arises because Electronic Health Record (EHR) systems are often designed to be generic, performing configurable tasks on various objects. In this example, if a caregiver finds the ECG result concerning, she will order a blood test (\texttt{ot}) as a standard care procedure, and the test result helps her to decide the next steps.
However, such a procedure is not visible in the OC-DFG due to the level of abstraction at which the data are processed.

To address this challenge, this paper introduces four operations (drill-down, roll-up, unfold, and fold) that enable users to dynamically adjust the level of detail in OCPM. These operations change the granularity of object types and event types. Together, these operations facilitate `zooming in and zooming out', enabling the discovery of process models at different levels of abstraction using current OCPM algorithms. The OC-DFG discovered by transforming the running example log using these operations is demonstrated in the center and right sections of \figurename~\ref{fig:runningexample}, showing how the standard care procedure can be revealed by setting the right level of abstraction.

The proposed operations are formally defined and implemented in an open-source Python library named \texttt{processmining}. Their effectiveness is evaluated through a case study on group-based student learning in a course at Stockholm University, representing the application of object-centric educational process mining. The dataset was collected from the learning management system, and it records how student groups progressed relative to predefined course milestones in the past four years. The impact of these operations is assessed by comparing the fitness and precision of Object-Centric Petri nets discovered before and after their application. The results demonstrate the ability to generate more accurate and representative process models. In rare cases where fitness did not improve, the logs were transformed into temporal Event Knowledge Graphs~\cite{khayatbashi2024transforming}, revealing issues related to rolling group membership. In addition, we assess the scalability of the operations on large, publicly available OCELs derived from the Business Process Intelligence Challenge datasets, showing that the approach remains computationally feasible on logs containing millions of events.

This paper extends our previous work~\cite{khayatbashi2025olap}, where we first introduced the four OLAP-inspired operations for Object-Centric Process Mining (OCPM) and provided an initial implementation. Beyond that initial concept, the present article:
\begin{itemize}
	\item provides complete formal definitions and algorithms for all four operations (drill-down, roll-up, unfold, and fold);
	\item conducts a comprehensive real-world case study on four years of educational process data, demonstrating significant improvements in process model quality, particularly in terms of \textit{precision} and \textit{fitness}, and uncovering meaningful behavioral patterns within evolving group structures;
	\item extends the analytical capabilities by integrating OCELs with temporal Event Knowledge Graphs (tEKGs), enabling the handling of dynamic object relationships (such as changes in group membership) over time and supporting richer process exploration; and
	\item evaluates the scalability of the proposed operations on large, publicly available OCELs derived from the Business Process Intelligence Challenge datasets, demonstrating their computational feasibility on industrial-scale event logs with millions of events.
\end{itemize}

The remainder of this paper is organized as follows: Section~\ref{sec:Background} summarizes the related background and further elaborates on the problem using the running example. Section~\ref{sec:Prelinimaries} provides the necessary preliminaries, while Section~\ref{sec:Approach} formally defines the proposed approach. Section~\ref{sec:Evaluation} presents the evaluation results and discussion. Finally, Section~\ref{sec:Conclusion} concludes the paper.


\section{Background}\label{sec:Background}
This section provides a brief summary of related work enabling analysts to adjust the level of analysis using traditional log formats in process mining. It then elaborates on the identified problem using the running example. 

\subsection{Related work}
In data analysis, the ability to drill down and roll up data is crucial for extracting meaningful insights from large datasets~\cite{van2013process}. This capability is particularly significant in multi-dimensional data analysis, where data is examined across various dimensions, adding complexity to the task. Tools like Microsoft Excel and Online Analytical Processing (OLAP) systems highlight the practical utility and importance of these techniques in real-world applications.

The concepts of drilling down and rolling up were recognized early in process mining~\cite{van2013process, van2014process}. van der Aalst introduced the concept of Process Cubes in 2013, emphasizing OLAP operations such as slice, dice, drill-down, and roll-up to support data-driven process analysis~\cite{van2013process}. Bolt and van der Aalst later implemented Process Cubes~\cite{van2013process, bolt2015multidimensional} as a ProM plugin and a standalone Java application. These implementations allowed analysts to apply OLAP operations to process cubes and transform the results into traditional event logs by mapping one of the attributes to the case ID. However, this work did not support multi-dimensional process mining because Object-Centric Process Mining (OCPM) had not yet been defined at that time.

Process cubes have since been applied in various domains. For example, Gupta and Sureka modeled a process cube with nine dimensions for defect resolution processes~\cite{gupta2014process}, demonstrating the application of OLAP operations. van der Aalst et al.~\cite{van2015comparative} used process cubes in education to compare the performance of student groups in a course. Bolt et al. proposed integrating process mining with analytic workflows for large-scale comparative analyses~\cite{bolt2015exploiting}. 
Jalali employed drill-down, roll-up, slice, and dice operations to investigate Dutch autonomous administrative authorities using process cubes, focusing on both control-flow and resource perspectives~\cite{jalali2016reflections}.

In healthcare, Weerdt et al. demonstrated how drill-up (another name for roll-up) and drill-down operators can reveal insights into care flows to improve clinical processes~\cite{de2013getting}. Additionally, Yeshchenko et al. highlighted the need for drill-down and roll-up operations in process drift analysis, extending the application of these techniques beyond process exploration to identifying concept drift~\cite{yeshchenko2021visual}.

Although slice-and-dice operations have been implemented through various filtering techniques in process mining tools, implementing drill-down and roll-up remains challenging. Analysts often perform these operations manually, which not only increases the risk of implementation errors during data cleaning and reshaping but also increases the risk of biased interpretations~\cite{van2018spreadsheets}.

The Object-Centric Event Log (OCEL) standard provides a systematic approach to defining such operations by establishing relationships between multiple objects and events. This creates a structured, multi-dimensional space where granularity levels can be adjusted across different components. In parallel with this paper, we have explored how drill-down operations can uncover more patterns using Markov-based clustering~\cite{Miri2024}. This work formally defines drill-down, roll-up, fold, and unfold operations, while also providing tools to apply them in practice, which is also demonstrated using a case study.

\subsection{Problem definition}
To illustrate the problem and elaborate on our approach, we extend the running example introduced in \figurename~\ref{fig:runningexample}.
We define a simple OCEL for the running example, summarized in \tablename~\ref{tbl:runninglog}. The table presents the data in two distinct views: \textit{Events with connected objects} and \textit{Objects with current attribute values}. It is important to emphasize that this running example is not intended to provide a detailed explanation of the OCEL 2.0 specification; for that, readers are referred to~\cite{berti2023ocelspecification}.

Each event in the log is characterized by an `Event ID', `Event Type', and `Timestamp', and is associated with a list of related objects (qualifiers are abstracted in this example). The `Related Objects' column contains object IDs that are detailed in a separate view. Each object is defined by an `Object ID', `Object Type', and its `Current Attribute Values'. For simplicity, this example does not depict how attribute values evolve over time or how relationships between objects are captured.

As can be seen in \tablename~\ref{tbl:runninglog}, the log records \emph{specific} test types such as \texttt{12-lead ECG}, \texttt{Monitoring ECG}, \texttt{Fasting Blood Glucose Test}, and \texttt{Arterial Blood Gas Test}. In contrast, caregivers typically reason in terms of broader categories, for example by talking about an `ECG test' or a `blood test'. In other words, domain experts are aware that different concrete investigations belong to higher-level conceptual groups (e.g., both \texttt{12-lead ECG} and \texttt{Monitoring ECG} are ECG tests, and both \texttt{Fasting Blood Glucose Test} and \texttt{Arterial Blood Gas Test} are blood tests), but these categories are not explicitly represented in the operational data. This situation is common when performing process mining on real-world information systems: the event log faithfully captures low-level details, while omitting some of the abstraction levels that users naturally employ when describing and interpreting the process. As a consequence, process discovery performed directly on such logs yields models at a fixed, often suboptimal, level of granularity, and important behavioral patterns that emerge only at coarser or alternative abstraction levels remain hidden.

\begin{table}[t]
	\centering
	\subfloat[\textbf{Events:} A view over events recorded with relations to multiple objects in an OCEL.\label{tbl:running-events}]{
		\begin{tabular}{llll}
			\toprule
			Event ID & Event Type (Activity) & Timestamp & Related Objects \\
			\midrule
			e1  & register patient (rp) & ts1 & [p1] \\
			e2  & order test (ot)       & ts2 & [p1, t1] \\
			e3  & register test (rt)    & ts3 & [p1, t1] \\
			e4  & order test (ot)       & ts4 & [p1, t2] \\
			e5  & register test (rt)    & ts5 & [p1, t2] \\
			e6  & register patient (rp) & ts6 & [p2] \\
			e7  & order test (ot)       & ts7 & [p2, t3] \\
			e8  & register test (rt)    & ts8 & [p2, t3] \\
			e9  & order test (ot)       & ts9 & [p2, t4] \\
			e10 & register test (rt)    & ts10 & [p2, t4] \\
		\end{tabular}
	}
	\hfill
	\subfloat[\textbf{Objects:} A view over objects with current attribute values in an OCEL.\label{tbl:running-objects}]{
		\begin{tabular}{lll}
			\toprule
			Object ID & Object Type & Current Attribute Values \\
			\midrule
			p1 & Patient & \{"name":"Jessica", ...\} \\
			p2 & Patient & \{"name":"Michael", ...\} \\
			t1 & Test    & \{"type":"12-lead ECG", ...\} \\
			t2 & Test    & \{"type":"Fasting Blood Glucose Test", ...\} \\
			t3 & Test    & \{"type":"Monitoring ECG", ...\} \\
			t4 & Test    & \{"type":"Arterial Blood Gas Test", ...\} \\
		\end{tabular}
	}
	\caption{An extended example OCEL log for the running example with two patients. Each patient undergoes
		a specific ECG test and a specific blood test, all recorded through the generic
		activities \texttt{order test} and \texttt{register test}.}
	\label{tbl:runninglog}
\end{table}

A process discovery algorithm can analyze sequences of events for each object type and identify the relationships among activities. For instance, considering object \texttt{p1}, which represents a \texttt{patient}, the following relationships can be observed:
\texttt{rp}${\color{darkblue}\stackrel[]{\texttt{p}}{\rightarrow}}$\texttt{ot}${\color{darkblue}\stackrel[\texttt{p}]{\texttt{p}}{\rightleftarrows}}$\texttt{rt}, where \texttt{{\color{darkblue}p}} represents \texttt{Patient}.
These relations can be observed by following the blue cubes in the upper-left side of \figurename~\ref{fig:runningexample}, showing the sequence of event types in relation to \texttt{Patient} object type that happened over time. 
For objects \texttt{t1} and \texttt{t2}, representing different \texttt{Tests}, the following relationships can be identified:
\texttt{ot}${\color{darkred}\stackrel[]{\texttt{t}}{\rightarrow}}$\texttt{rt}, where \texttt{{\color{darkred}\texttt{t}}} represents \texttt{Test}.

The overall Object-Centric Directly-Follows Graph (OC-DFG) is constructed as the union of all these relationships, as shown on the left side of \figurename~\ref{fig:runningexample}. However, this abstraction fails to reveal direct relationships between ordering different tests. This limitation arises because the algorithm abstracts the logs at the object type level, in this case, \texttt{Test}, and does not distinguish activities by the specific objects they refer to.

To address these challenges, we introduce four OLAP-inspired operations on OCELs.
\textit{Drill-down} refines object types; for example, a generic \texttt{Test} object can be split into objects representing individual investigations such as 12-lead ECG, monitoring ECG, fasting blood glucose, and arterial blood gas. 
\textit{Roll-up} increases the level of aggregation for object-types: it can merge several concrete tests into categories that caregivers use in practice, for instance merging the drilled-down objects \texttt{12-lead ECG} and \texttt{Monitoring ECG} into an \texttt{ECG Test} category. 
\textit{Unfold} refines event types with respect to these object types, so that generic activities like \texttt{order test} become variants such as \texttt{order ECG Test}. 
\textit{Fold} collapses these variants back into the original generic event types. Together, these operations enable a more nuanced exploration of object-centric event logs by allowing users to dynamically adjust the level of detail based on the abstractions relevant to their analysis.

\subsection{Implications for Event Knowledge Graphs}
Object-centric event data do not need to be represented only as OCELs. Graph-based representations, such as Event Knowledge Graphs (EKGs)~\cite{esser2021multi} and their temporal extension, temporal Event Knowledge Graphs (tEKGs)~\cite{khayatbashi2024transforming}, provide an alternative representation of OCED in which events, objects, attributes, and their temporal relations are modeled as a labeled property graph. In a tEKG, events, objects, and object snapshots over time are graph nodes, while relations such as event-object and object-object relations become edges enriched with temporal and semantic properties.

This property-graph perspective differs from traditional relational or table-based representations. In a tEKG, relationships are first-class citizens: they can be queried and traversed directly using graph queries, and exploited by a rich toolbox of graph algorithms. For example, centrality measures (e.g., degree, betweenness, or eigenvector centrality) can be used to identify influential objects (such as highly connected resources or `hub' groups), community detection and clustering can reveal structurally coherent sub-processes, or cohorts of objects. 

Because relationships are stored as explicit edges, multi-hop patterns (e.g., chains of object-to-object links or evolving group membership) are more naturally expressed and efficiently traversed in graph models than in purely relational schemas, where equivalent queries must be reconstructed via multiple complex joins, as relational databases are not designed for traversal-centric workloads~\cite{vicknair2010comparison}. Moreover, temporal information can be attached directly to nodes and edges, enabling the use of temporal path queries and time-aware variants of these algorithms to reason about how object relations and interaction structures change over time~\cite{khayatbashi2024transforming}.

In this work, the four OLAP-inspired operations (drill-down, roll-up, unfold, and fold) are formally defined and implemented on top of OCEL 2.0. However, they are not tied to OCEL as a storage or execution format: earlier work~\cite{khayatbashi2023transforming,khayatbashi2024transforming} has introduced bidirectional transformation techniques between OCEL and EKG/tEKG. As a consequence, the same OLAP operations can be applied to any OCED representation that can be transformed to OCEL, including tEKG, by (i) transforming the graph to OCEL, (ii) applying the operations on the OCEL, and if needed (iii) transforming the resulting log back into a tEKG. This enables analysts to combine OCEL-based OLAP operations with graph-based querying, graph algorithms, and visualization in a complementary way. As our main focus is not on tEKG, we refer readers to~\cite{khayatbashi2024transforming} for further details.


\section{Preliminaries}\label{sec:Prelinimaries}

This section provides a summary of the definition of OCEL, as adopted from~\cite{berti2023ocelspecification,khayatbashi2024transforming}. This definition serves as the foundation for formalizing multi-dimensional data operations in the subsequent sections.
We begin by defining the universes upon which the formal definition of OCEL 2.0 is defined.

\vspace{0.4cm}

\begin{definition}\label{def:univ}\normalfont 
	We assume the existence of these \textbf{universes}~\cite{berti2023ocelspecification,khayatbashi2024transforming}:
\end{definition}

\begin{itemize}
	\item $\univ{eid}$ is the universe of event identifiers,
	\item $\univ{\mathit{att}}$ is the universe of attribute names,,
	\item $\univ{oid}$ is the universe of object identifiers,
	\item $\univ{\mathit{val}}$ is the universe of attribute values
	\item $\univ{\mathit{etype}}$ is the universe of event types,
	\item $\univ{\mathit{time}}$ is the universe of timestamps, and 
	\item $\univ{\mathit{otype}}$ is the universe of object types,
	\item $\univ{\mathit{qual}}$ is the universe of qualifiers.
\end{itemize}

In our running example, 
$\univ{eid}=\{ \mathtt{e1}, \mathtt{e2}, \ldots, \mathtt{e10} \}$, 
$\univ{oid}=\{ \mathtt{p1}, \mathtt{p2}, \mathtt{t1}, \ldots, \mathtt{t4} \}$, 
$\univ{\mathit{etype}}=\{\mathtt{rp},\ \mathtt{ot},\ \mathtt{rt} \}$,
$\univ{\mathit{otype}}=\{ \mathtt{Patient}, \mathtt{Test} \}$, 
$\univ{\mathit{att}}=\{ \mathtt{name}, \mathtt{type},  \ldots  \}$, 
$\univ{\mathit{time}}=\{ \mathtt{ts1}, \ldots, \mathtt{ts10}\}$,
$\univ{\mathit{val}}=\{ \mathtt{Jessica}, \mathtt{12-lead ECG},  \mathtt{Fasting Blood Glucose Test},  \ldots \}$.

\begin{definition}\label{def:ocel}
	\normalfont
	An \textbf{Object-Centric Event Log~(OCEL)}~$L$ is a tuple~$(E,$  $O,\mathit{EA},$ $\mathit{OA},$ $\mathit{evtype},\mathit{evid},\mathit{time},\mathit{objtype},\mathit{objid},\mathit{eatype},\mathit{oatype},\mathit{eaval}, $ 
	$\mathit{oaval},\mathit{E2O},\mathit{O2O},\mathit{OT},\mathit{ET})$
	where~\cite{berti2023ocelspecification,khayatbashi2024transforming}:
	\begin{itemize}
		\item $E$ and $O$ are sets of events and sets of objects, where $E\cap O = \emptyset$,
		\item $\mathit{EA} \!\subseteq\! \univ{\mathit{att}}$ and $\mathit{OA} \!\subseteq\! \univ{\mathit{att}}$ are sets of
		attributes for events and objects, respectively,
        \item $ET$ and $OT$ are sets of dynamic event types and object types used in the log,
		\item $\mathit{evtype}: E \rightarrow \mathit{ET}$ is a function assigning event types to events,
		\item $\mathit{evid}: E \rightarrow \univ{\mathit{eid}}$ is a function assigning event id to events,
		\item $\mathit{time}: E \rightarrow \univ{\mathit{time}}$ is a function assigning timestamps to events, 
		\item $\mathit{objtype}: O \rightarrow \mathit{OT}$ is a function assigning object types to objects,
		\item $\mathit{objid}: O \rightarrow \univ{\mathit{oid}}$ is a function assigning object id to objects,
		\item $\mathit{eatype}: \mathit{EA} \rightarrow \mathit{ET}$ is a function assigning event types to event attributes, 
		\item $\mathit{oatype}: \mathit{OA} \rightarrow \mathit{OT}$ is a function assigning object types to object attributes, 
		\item $\mathit{eaval}: (E \times \mathit{EA}) \not\rightarrow \univ{\mathit{val}}$ is a partial function assigning values to (some) event attributes such that $evtype(e) = eatype(ea)$ for all $(e,ea)\in dom(eaval)$,
		\item $oaval: (O \times \mathit{OA} \times \univ{\mathit{time}}) \not\rightarrow \univ{\mathit{val}}$ assigns values to object attributes such that $objtype(o)=oatype(oa)$ for all $(o,oa,t)\in dom(oaval)$,
		\item $\mathit{E2O}\subseteq E \times \univ{\mathit{qual}} \times O$ are the qualified event-to-object relations, and
		\item $\mathit{O2O}\subseteq O \times \univ{\mathit{qual}} \times O$ are the qualified object-to-object relations.

	\end{itemize}
\end{definition}

For any partial function $f: D\not\rightarrow R$, if there exists $d\in D$ where $d\notin dom(f)$, we say $f(d)=\perp$. 
We gave an excerpt of the OCEL 2.0 definition that we needed in this paper. We refer the readers to~\cite{berti2023ocelspecification} for the full specification.


\section{Approach}\label{sec:Approach}
\begin{figure}[h!]
	\centering
	\includegraphics[width=\linewidth]{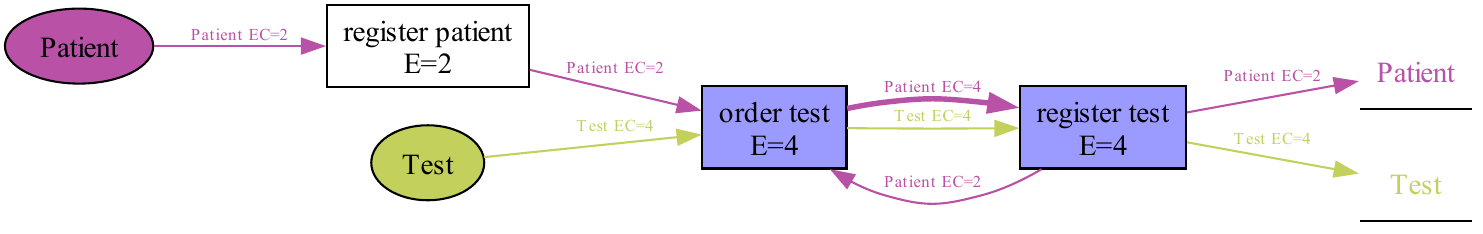}
	\caption{OC-DFG of the running example before applying any operations}
	\label{fig:DFG}
\end{figure}

This section elaborates on the proposed solution for enabling OLAP operations on Object-Centric Event Logs (OCELs), namely drill-down, roll-up, unfold, and fold. 
Each operation is defined both formally and informally, supported by a running example to facilitate understanding.
We begin by presenting the Object-Centric Directly-Follows Graph (OC-DFG) of the running example prior to applying any OLAP operations, as shown in \figurename~\ref{fig:DFG}, which serves as a baseline for comparison. For each operation, a formal definition specifies the underlying transformation, while an informal explanation provides intuition, followed by an illustration of its effect on the running example.
The outcomes of applying these operations are demonstrated in this section and are also available in the accompanying materials\footnote{See \url{https://github.com/shahrzadkhayatbashi/olap-operations4ocel}}.
The section concludes with a brief discussion of the proof-of-concept implementation, which makes these operations accessible for research and experimentation.

\subsection{Drill-down}

We refine object types by representing them as tuples consisting of the original object type and a selected attribute value (e.g., the value of the \texttt{type} attribute). This enables a more fine-grained view by explicitly distinguishing variants that are otherwise aggregated. The transformation can be applied recursively to achieve deeper levels of differentiation.

\begin{figure}[t!]
	\centering
	\includegraphics[width=\linewidth]{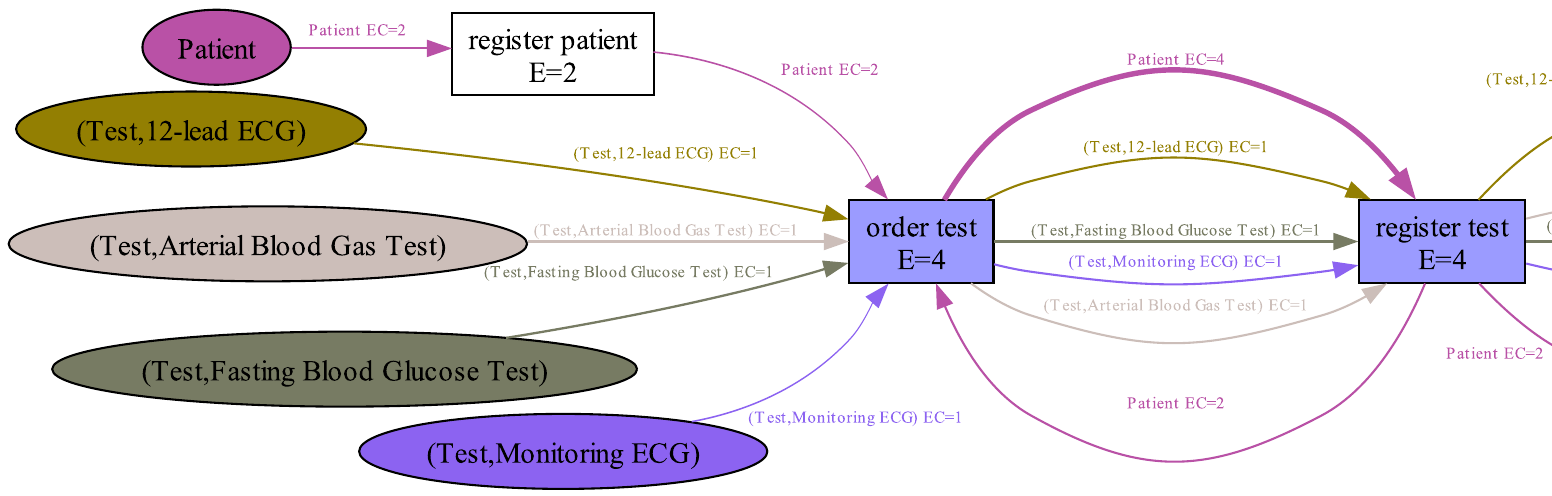}
	\caption{OC-DFG of the running example after applying drill-down operation on Test object type}
	\label{fig:DFG_drilled}
\end{figure}

In the running example, the object type \texttt{Test} is replaced by tuples combining \texttt{Test} with the corresponding \texttt{type} attribute value. As a result, object types such as \texttt{(Test, 12-lead ECG)}, \texttt{(Test, Fasting Blood Glucose Test)}, \texttt{(Test, Monitoring ECG)}, and \texttt{(Test, Arterial Blood Gas Test)} are obtained. This allows the discovery of OC-DFGs that distinguish between different test types, as illustrated in \figurename~\ref{fig:DFG_drilled}.

\begin{algorithm}[t!]
	\caption{Drillling-down OCEL based on an Object Type and Attribute}\label{alg:drilldown}
	\SetKwProg{Fn}{Function}{:}{}
	\SetKwFunction{FDrilldown}{drill-down}
	\Fn{\FDrilldown} {
		\KwData{
			$\big(L=(E, O, \mathit{EA}, \mathit{OA}, \mathit{evtype}, \mathit{evid}, \mathit{time}, \mathit{objtype}, \mathit{objid}, \mathit{eatype}, $ 
			$\mathit{oatype}, \mathit{eaval},  $ 
			$ \mathit{oaval}, \mathit{E2O}, $ 
			$\mathit{O2O},\mathit{OT},\mathit{ET}), ot\in \univ{\mathit{otype}}$, $oa\in OA\big)$
		}
		
		\KwResult{$L$, drilled-down OCEL}
		\ForEach{\normalfont \label{alg1:foreach:start}
			$(o,\mathit{oa},t)\in dom(\mathit{oaval})$} 
		{%
			$\mathit{val} \leftarrow \mathit{oaval}(o,\mathit{oa},t)$\\
			\If{$(\mathit{val}\neq\perp) \wedge (\mathit{objtype}(o)=\mathit{ot})$}{ \label{alg1:if:start}
				\tcp{extending object types with object attribute values}
				$\mathit{OT} \leftarrow \mathit{OT} \cup \{(\mathit{ot},\mathit{val})\}$ \\ 
				\label{alg1:extend:universe}
				\tcp{drilling-down object types}
				Modify $\mathit{objtype}$ such that $\mathit{objtype}(o)=(\mathit{ot},\mathit{val})$\\     \label{alg1:modify:objtype}
				\tcp{drilling-down object attributes types}
				Modify $\mathit{oatype}$ such that $\mathit{oatype}(\mathit{oa})=(\mathit{ot},\mathit{val})$\\   
				\label{alg1:modify:oatype}
			}
		}   
		\KwRet{$L$}\;\label{alg1:line:return}
	}
\end{algorithm}

Algorithm~\ref{alg:drilldown} formalizes this transformation. It iterates over all values of the selected attribute for objects of the given type (line~\ref{alg1:foreach:start}). For example, in the running example, $\mathtt{12-lead ECG}$ is the value of the $\mathtt{type}$ attribute for the $\mathtt{Test}$ object with identifier $\mathtt{t1}$. If a value is defined for that object type (line~\ref{alg1:if:start}), the algorithm extends the object types of the log ($\mathit{OT}$) with a new drilled member, such as $\texttt{(Test, 12-lead ECG)}$ (line~\ref{alg1:extend:universe}). It then modifies the object type of the selected objects to the drilled version (line~\ref{alg1:modify:objtype}) and applies the same changes to the object attribute types (line~\ref{alg1:modify:oatype}). This ensures that both object types and object attribute types are drilled down simultaneously. Finally, the algorithm returns the modified log as output (line~\ref{alg1:line:return}).
It is worth noting that the drill-down operation can be performed multiple times in sequence. The drilled-down version of the log enables the analysis of the process at a finer level of granularity - distinguishing the types of tests.

In this drilled-down view, each concrete investigation (12-lead ECG, monitoring ECG, fasting blood glucose, arterial blood gas) is treated as a separate object type, and the corresponding OC-DFG in \figurename~\ref{fig:DFG_drilled} reveals more fine-grained behavioral patterns than the original, highly aggregated model.

However, the level of abstraction provided by this drill-down is still not fully aligned with how caregivers conceptualize the process. As discussed earlier, clinicians typically reason in terms of broader categories such as ECG tests and blood tests, rather than individual test names stored in the operational system. The running example OCEL, in contrast, records only detailed labels and does not explicitly encode these higher-level categories. As a result, the drilled-down OC-DFG may be overly fine-grained from the analyst's perspective, thereby obscuring more general patterns that hold across variants of the same clinical concept.

In our implementation, we support two variants of the drill-down operation with respect to object attributes whose values may change over time. The formalization in Algorithm~\ref{alg:drilldown} is based on the temporal object attribute function \(\mathit{oaval}\) and therefore takes \emph{historical} values into account: for each triple \((o,\mathit{oa},t)\) it uses the value \(\mathit{oaval}(o,\mathit{oa},t)\), so that the same object can be associated with different drilled-down types at different time points if its attribute value changes. This realizes a fully time-aware drill-down where object types can evolve along the execution of the process. 

In the library, we additionally provide a simplified, state-based variant that ignores historical changes and only considers the \emph{current} value of the selected attribute for each object (e.g., the value at the last timestamp or as stored in the OCEL without history). This second variant can be seen as applying Algorithm~\ref{alg:drilldown} on a conceptually pre-filtered log in which \(\mathit{oaval}\) contains exactly one value per object and attribute. Since it is a special case of the temporal formulation (obtained by discarding all but the current value), we do not introduce a separate formalization for it.

To obtain process models at a level of abstraction that better matches the caregivers' mental model, a complementary operation is required to \emph{roll up} drilled-down object types into clinically meaningful groups, such as aggregating 12-lead ECG and monitoring ECG into an \texttt{ECG Test} category, and fasting blood glucose and arterial blood gas into a \texttt{Blood Test} category. The following subsections elaborate on the roll-up operation.

\subsection{Roll-up}

We propose aggregating object types into higher-level categories by grouping multiple refined object types under a common abstraction. Formally, this operation defines a mapping from a set of detailed object types to a more general object type. Informally, roll-up provides a coarser view of the process by merging variants that belong to the same conceptual category.

\begin{figure}[t!]
	\centering
	\includegraphics[width=\linewidth]{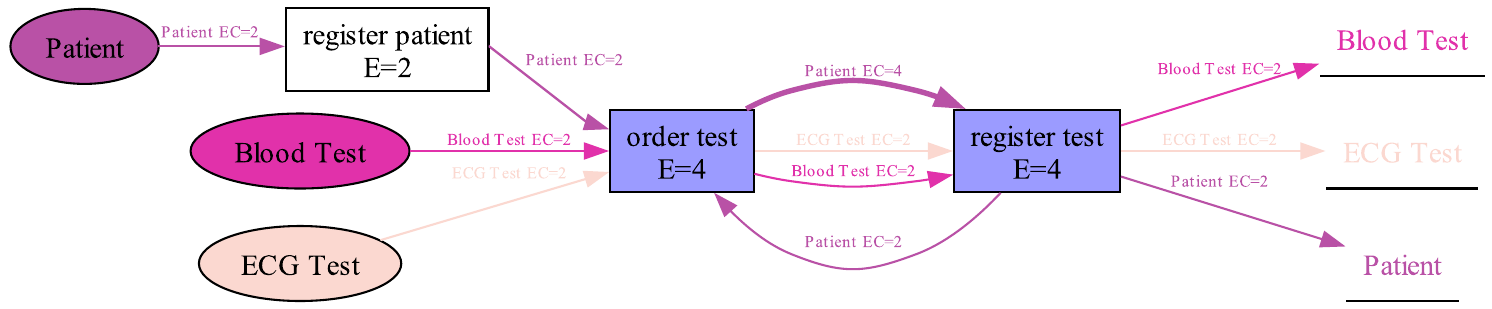}
	\caption{OC-DFG of the running example after applying roll-up operation}
	\label{fig:DFG_rollup}
\end{figure}

In the running example, the drilled-down object types \texttt{(Test, 12-lead ECG)} and \texttt{(Test, Monitoring ECG)} are aggregated into the higher-level category \texttt{ECG Test}, while \texttt{(Test, Fasting Blood Glucose Test)} and \texttt{(Test, Arterial Blood Gas Test)} are grouped into \texttt{Blood Test}. This transformation increases the abstraction level and enables the discovery of OC-DFGs that reflect clinically meaningful groupings as shown in \figurename~\ref{fig:DFG_rollup}.

\begin{algorithm}[t!]
	\caption{Rolling-up OCEL based on a set of Object Types}\label{alg:rollup}
	\SetKwProg{Fn}{Function}{:}{}
	\SetKwFunction{FRollup}{roll-up}
	\Fn{\FRollup} {
		\KwData{
			$\big(L=(E, O, \mathit{EA}, \mathit{OA}, \mathit{evtype}, \mathit{evid}, \mathit{time}, \mathit{objtype}, \mathit{objid}, \mathit{eatype}, $ 
			$\mathit{oatype}, \mathit{eaval},  $  
			$ \mathit{oaval}, \mathit{E2O}, $ 
			$\mathit{O2O},\mathit{OT},\mathit{ET}), \mathit{OT}^\prime\subseteq \mathit{OT}, \mathit{ot}^\prime\in\univ{\mathit{otype}}\big)$
		}
		\KwResult{$L$, rolled-up OCEL}
		
		\tcp{extending the object types of the log}
		$\mathit{OT} \leftarrow \mathit{OT} \cup \{\mathit{ot}^\prime\}$\\     \label{alg2:extendOT}
		\ForEach{\normalfont \label{alg2:iterate_ots} 
			$o\in O$, where $\mathit{objtype}(o)\in\mathit{OT}^\prime$}   
		{%
			\tcp{rolling-up object types}
			Modify $\mathit{objtype}$ such that $\mathit{objtype}(o)=\mathit{ot}^\prime$\\     \label{alg2:line:replacingobjectsot}
			\ForEach{\normalfont \label{alg2:foreach:start2}  \label{alg2:for_oas}
				$(o, \mathit{oa},t)\in dom(\mathit{oaval})$}   
			{
				\tcp{rolling-up object attributes types}
				Modify $\mathit{oatype}$ such that $\mathit{oatype}(\mathit{oa})=\mathit{ot}^\prime$\\   \label{alg2:line:replacingoaot}
			}
		} 
		\KwRet{$L$}\;
	}
\end{algorithm}

Algorithm~\ref{alg:rollup} outlines the procedure for rolling up an OCEL $L$ using a specified set of fine-grained object types $\mathit{OT}^\prime$, which are to be transformed into a coarser-grained object type $\mathit{ot}^\prime$.
In our example, we may need to apply it twice: first specifying $\mathit{OT}^\prime=\{\texttt{(Test, 12-lead ECG)}, \texttt{(Test, Monitoring ECG)}\}$ and the coarse-grained object type as \texttt{ECG Test}, and second specifying $\mathit{OT}^\prime=\{\texttt{(Test, Fasting Blood Glucose Test)}, \texttt{(Test, Arterial Blood Gas Test)}\}$ and the coarse-grained object type as \texttt{Blood Test}.

Let us consider rolling up the log for \texttt{ECG Test}. The algorithm starts by extending the log's set of object types to include the new coarse-grained type (line~\ref{alg2:extendOT}), e.g., \texttt{ECG Test}. Then, it iterates over all objects of the specified fine-grained object types and increases their level of abstraction by assigning the coarser type to them (line~\ref{alg2:line:replacingobjectsot}), and applies the same changes to all their object attributes (line~\ref{alg2:line:replacingoaot}). Finally, it returns the rolled-up OCEL.

The roll-up operation can also be performed multiple times in sequence. The OC-DFG shown in \figurename~\ref{fig:DFG_rollup} is obtained by applying the roll-up procedure twice to the running example logs, as described above.

\subsection{Unfold}

While roll-up and drill-down operate on the abstraction level of object types, the \emph{unfold} operation focuses on the structural relationships between events and objects. Specifically, unfolding makes implicit event-object associations explicit by restructuring the OCEL such that each event-object relation is represented in a more fine-grained and analyzable form. This is achieved by projecting the event type to a combination of the event type and the object type, thereby segregating activities based on specific object types. For instance, unfolding the \texttt{ot} and \texttt{rt} activities with rolled-up object types in the running example changes the event type of \texttt{e2} from \texttt{ot} to \texttt{(ot, ECG Test)}.

\begin{figure}[b!]
	\centering
	\includegraphics[width=\linewidth]{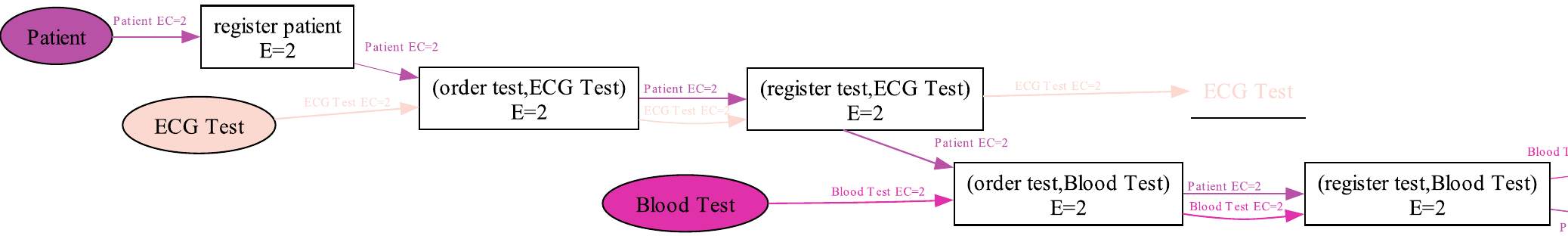}
	\caption{OC-DFG of the running example after applying unfold operation}
	\label{fig:DFG_unfold}
\end{figure}

If we apply unfolding in the running example to all events related to different tests, we obtain the OC-DFG in \figurename~\ref{fig:DFG_unfold}, which highlights object-specific behavior more transparently. In particular, it reveals that ECG tests are always performed before blood tests - an ordering constraint that was hidden in the original OC-DFG.

\begin{algorithm}[t!]
	\caption{Unfolding OCEL based on an Event Type and Object Type}\label{alg3:unfold}
	\SetKwProg{Fn}{Function}{:}{}
	\SetKwFunction{FUnfold}{unfold}
	\Fn{\FUnfold} {
		\KwData{
			$\big(L=(E, O, \mathit{EA}, \mathit{OA}, \mathit{evtype}, \mathit{evid}, \mathit{time}, \mathit{objtype}, \mathit{objid}, \mathit{eatype}, $ 
			$\mathit{oatype}, \mathit{eaval}, \mathit{oaval}, \mathit{E2O}, $ 
			$\mathit{O2O},\mathit{OT},\mathit{ET}),$ 
			$\mathit{et}\in \mathit{ET}$, 
			$ot\in \mathit{OT}$,
			$Q\subset\univ{\mathit{qual}}\big)$
		}
		\KwResult{$L$, unfolded OCEL}
		
		\tcp{filtering unfoldable events}
		$\mathit{UE} \leftarrow \{(e,q,o)\in\mathit{E2O}\ |\ \mathit{et}=\mathit{evtype}(e) \wedge  q\in Q \wedge \mathit{otype}(o)=\mathit{ot}\}$ \\ \label{alg3:unfoldableevents}
		\label{alg3:unfoldable:events}
		
		\ForEach{\normalfont \label{alg3:foreach:start}
			$(e,q,o)\in\mathit{UE}$} 
		{%
			
			\tcp{extending event types with unfolded event type}
			$\mathit{ET} \leftarrow \mathit{ET} \cup \{(\mathit{et},\mathit{ot})\}$ \\ \label{alg3:extend:type:universe}
			\tcp{unfold event types for events}
			Modify $\mathit{evtype}$ such that $\mathit{evtype}(e)=(\mathit{et},\mathit{ot})$\\     \label{alg3:changeeventtype}
			\ForEach{\normalfont \label{alg3:foreach:inner:start} \label{alg3:startfor}
				$(e,ea)\in dom(\mathit{eaval})$} 
			{%
				\If{$(\mathit{eaval}(e,\mathit{ea})\neq\perp) \wedge (\mathit{evtype}(e)=\mathit{et})$}{ 
					\tcp{unfold event types for events attributes}
					Modify $\mathit{eatype}$ such that $\mathit{eatype}(\mathit{ea})=(\mathit{et},\mathit{ot})$\\   \label{alg3:modifyingea}
				}
			}
		}
		\KwRet{$L$}\;
	}
\end{algorithm}

Algorithm~\ref{alg3:unfold} describes the process of unfolding the OCEL $L$ based on a given event type $\mathit{et}$ and object type $\mathit{ot}$, using a qualifier from a given set of qualifiers  $Q$.
The algorithm begins by filtering all event-to-object relations where the qualifiers are within the desired list (line~\ref{alg3:unfoldableevents}). For each of these relations, it extends the event types by creating a tuple of the event type and the selected object type. For example, unfolding events with the type of \texttt{order test} over \texttt{ECG Test} extends the event types with $\texttt{\big(order test, ECG Test\big)}$ (line~\ref{alg3:extend:type:universe}). The algorithm then modifies the event type by including the tuple of the event type and object type (line~\ref{alg3:changeeventtype}). Subsequently, the unfolding algorithm updates the event type of each event attribute to reflect the tuple of the event type and object type (lines~\ref{alg3:startfor}--\ref{alg3:modifyingea}). Finally, it returns the unfolded event log.

In practice, events may be related to multiple objects, potentially of several object types that satisfy the unfolding criterion. Algorithm~\ref{alg3:unfold} operates on one object type \(\mathit{ot}\) at a time. If a single event is linked to multiple objects of that same type (e.g., an \texttt{order test} event referring to two different \texttt{ECG Test} objects), the event type will be changed based on the selected candidate. Please note that all existing event--object relations remain intact in the log. Thus, the structural connectivity of the OCEL is preserved and the unfolded event still participates in all its original relations. 

When an event is related to objects of several different types that should all be unfolded (e.g., both \texttt{ECG Test} and \texttt{Blood Test}), unfolding is applied separately for each object type. This can be done sequentially: we first call the unfold operation with \(\mathit{ot} = \texttt{ECG Test}\), then with \(\mathit{ot} = \texttt{Blood Test}\), etc. Each call extends the event type alphabet by a new pair \((\mathit{et},\mathit{ot})\) and updates the type of all matching events accordingly.

\subsection{Folding}

Folding increases the level of abstraction for event types and acts as the inverse of unfolding. Unfolding refines the event perspective by splitting event types according to the object types they refer to, whereas folding restores a log to a coarser representation by merging selected event types into a higher-level event type.

Conceptually, folding operates on the level of event types rather than object types. It takes selected event types and collapses them into a coarser-level type. In doing so, it recombines more fine-grained variants of the same activity into a single, more general activity label, while preserving the underlying event-object relations in the OCEL structure.

We can apply this operation to reverse the transformation carried out in the previous step by folding \texttt{(order test, ECG Test)} and \texttt{(order test, Blood Test)} into \texttt{order test}, and similarly \texttt{(register test, ECG Test)} and \texttt{(register test, Blood Test)} into \texttt{register test}. The folded log will lead to the discovery of an OC-DFG equivalent to the one in \figurename~\ref{fig:DFG_rollup}, and if we also roll up \texttt{ECG Test} and \texttt{Blood Test} to \texttt{Test}, we obtain the original OC-DFG in \figurename~\ref{fig:DFG}.

\begin{figure}[t!]
	\centering
	\includegraphics[width=\linewidth]{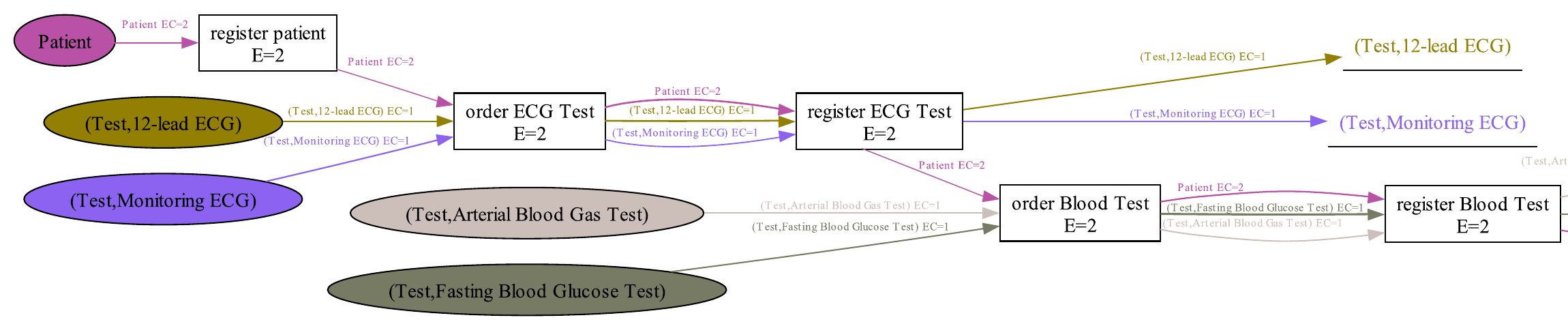}
	\caption{OC-DFG of the running example showing abstracting event types by folding}
	\label{fig:DFG_fold2}
\end{figure}

Please note that we can first drill down the original OC-DFG and unfold all event types to obtain a very fine-grained model. Next, we can fold all variations of ECG test events into `register ECG Test' and, analogously, fold all variations of blood test events into  `register Blood Test'~\footnote{These steps are illustrated in the accompanying running example on GitHub, which makes it easier to follow the text.}. The resulting OC-DFG is shown in \figurename~\ref{fig:DFG_fold2}.

Subsequently, we can roll up different tests to obtain the OC-DFG depicted in \figurename~\ref{fig:DFG_unfold}. This demonstrates that abstractions based on folding and roll-up can be flexibly applied according to the analyst's needs, enabling process models to be represented at higher levels of abstraction.

Algorithm~\ref{alg4:fold} describes the process of folding the OCEL $L$ based on a given set of event types $\mathit{ET}^\prime$ to $\mathit{et}^\prime$. This algorithm follows a straightforward process, where it changes the event type of all events whose current type is in the given set (see lines~\ref{alg4:1stfor}--\ref{alg4:changeevtype}). It then applies the same process to each event attribute type (see lines~\ref{alg4:foreach2:start}--\ref{alg4:changeea}). Finally, the algorithm returns the folded log.

\begin{algorithm}[t!]
	\caption{Folding OCEL based on a set of Event Types}\label{alg4:fold}
	\SetKwProg{Fn}{Function}{:}{}
	\SetKwFunction{FFold}{fold}
	\Fn{\FFold} {
		\KwData{
			$\big(L=(E, O, \mathit{EA}, \mathit{OA}, \mathit{evtype}, \mathit{evid}, \mathit{time}, \mathit{objtype}, \mathit{objid}, \mathit{eatype}, $ 
			$\mathit{oatype}, \mathit{eaval}, $
			$ \mathit{oaval}, \mathit{E2O}, $ 
			$\mathit{O2O},\mathit{OT},\mathit{ET}), \mathit{ET}^\prime\subseteq \mathit{ET}$, 
			$\mathit{et}^\prime\in \univ{\mathit{etype}}\big)$
			
		}
		\KwResult{$L$, folded OCEL}
		\tcp{extending event types with folded event type}
		$\mathit{ET} \leftarrow \mathit{ET} \cup \{\mathit{et}^\prime\}$ \\ \label{alg4:extend:type:universe}
		\ForEach{\normalfont \label{alg4:foreach:start}  \label{alg4:1stfor}
			$e\in E$, where $\mathit{evtype}(e)\in\mathit{ET}^\prime$} 
		{%
			Modify $\mathit{evtype}$ such that $\mathit{evtype}(e)=\mathit{et}^\prime$\\     \label{alg4:changeevtype}
		}
		\ForEach{\normalfont \label{alg4:foreach2:start} \label{alg4:2ndfor}
			$ea\in EA$, where $\mathit{eatype}(ea)\in\mathit{ET}^\prime$} 
		{%
			Modify $\mathit{eatype}$ such that $\mathit{eatype}(ea)=\mathit{et}^\prime$\\      \label{alg4:changeea}
		}
		\KwRet{$L$}\;
	}
\end{algorithm}

\subsection{Tools support}

We have implemented our approach as a proof of concept, providing the algorithms in an open-source Python library named \texttt{processmining}~\footnote{The library can be installed using \texttt{!pip install processmining}}. To facilitate reproducibility and broader application, the running example OCEL and the accompanying code demonstrating the application of these operations are made available on GitHub~\footnote{\url{https://github.com/shahrzadkhayatbashi/olap-operations4ocel}}. This allows readers to both replicate the results presented for the running example and apply the operations to their own object-centric event logs.

\section{Evaluation and Discussion}\label{sec:Evaluation}
The proposed operations have already been applied in both the educational~\cite{ocpm22025} and insurance~\cite{khayatbashiAI2025} domains, demonstrating their practical usefulness and versatility. Using the same set of operations in two distinct contexts not only illustrates their effectiveness in real-world scenarios, but also provides a form of methodological triangulation. This cross-domain application strengthens the robustness of our findings, as it shows that the benefits of the artefact are not limited to a single setting and can be generalized across domains. It also highlights the adaptability of the approach to domain-specific challenges, reinforcing its value as a flexible tool for object-centric process mining.

Nevertheless, these applications were not designed as systematic evaluations of the operations themselves. To address this gap, we conduct an evaluation along two complementary dimensions. First, we apply the operations in a real educational case study to illustrate and assess their effectiveness in practice. Second, we apply them to publicly available datasets to evaluate their performance and scalability when dealing with large OCELs.

\subsection{Case Study: Educational Domain}
To assess the effectiveness of the proposed OLAP operations in a realistic setting, we first applied them in the educational domain. In the following, we elaborate on the data extraction process and how these operations were applied, which enabled the discovery of behavioral patterns that were not visible at the original level of granularity. We also demonstrate how the application of these operations supports a deeper analysis of the assignment submission and improving the grading process. The empirical evaluation presented in this section is limited to the \textit{drill-down} and \textit{unfold} operations, as the objectives of the case study concerned the exploration of finer-grained behavioral patterns. Consequently, the inverse operations, \textit{roll-up} and \textit{fold}, were not evaluated independently in this case study because they were not required to achieve the analytical objectives; when applied to reverse a previous \textit{drill-down} or \textit{unfold} step, they exhibit equivalent computational behavior.

\subsubsection{Log Extraction}
We evaluated our proposed approach by applying our implementation to analyze real-world Object-Centric Event Logs (OCELs) extracted from educational processes following the OCPM$^2$ methodology~\cite{ocpm22025} using PM4Moodle~\cite{miri2026object}. The dataset, spanning four consecutive years, was sourced from the Learning Management System (LMS) and pertained to a Business Process Management (BPM) course offered by the Department of Computer and Systems Sciences at Stockholm University. This course incorporated diverse BPM activities, such as process modeling, analysis, and mining, delivered through group work and experiential learning~\cite{jalali2018teaching}. We focused on extracting the `submit assignment' and `set grade' events to evaluate how the process can be analyzed based on different submissions students made during the course.

To ensure anonymity, we removed any information that could potentially identify students, such as personal identification numbers, IP addresses, and other sensitive data. The data was then transformed into OCEL 2.0, incorporating extensions to capture dynamic changes in object-to-object relationships over time.
An example of such dynamic relationships is the connection between students and groups, which can evolve as students switch groups during the course. The current OCEL 2.0 standard and its implementation do not directly support this scenario. To address this limitation, we introduced qualifiers to associate students with both their former and the last valid groups (we call them the current group) and recorded valid period timestamps for these relationships. This enhancement allows us to filter and analyze both current and past group associations using existing implementations. Furthermore, the OCEL logs were transformed into a temporal Event Knowledge Graph (tEKG)~\cite{khayatbashi2024transforming}, enabling advanced querying and detailed case analysis.

\subsubsection{Data summary}
\begin{figure}[h]
	\centering  
	\includegraphics[width=0.32\textwidth]{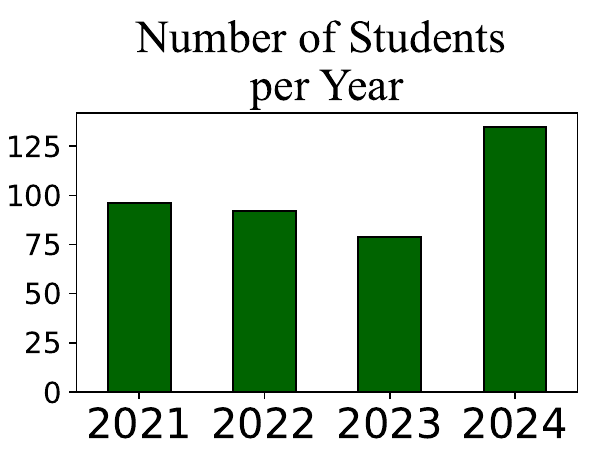}  
	\includegraphics[width=0.32\textwidth]{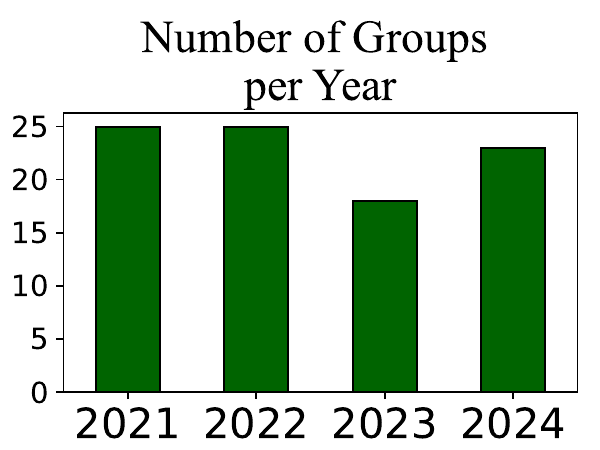}  
	\includegraphics[width=0.32\textwidth]{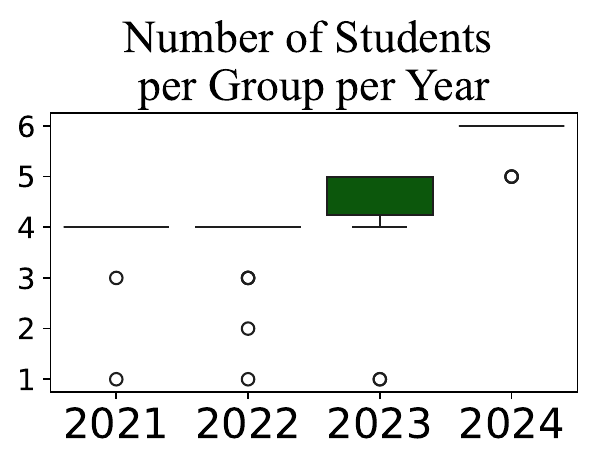}  
	\caption{Overview of student and group distributions}
	\label{fig:overall_case}
\end{figure}

The extracted OCEL files contain data for 401 students registered across 91 groups over four academic years (2021-2024). \figurename~\ref{fig:overall_case} illustrates, from left to right, the distribution of students per year, the number of groups per year, and the number of students per group per year. In 2021 and 2022, most groups consisted of four students, while in later years the group size increased. Although the majority of groups followed the standard structure defined by the course design, there were notable exceptions. In particular, three single-student groups were created to accommodate individual study arrangements, such as PhD students or students requiring independent examination.

The scope of the dataset is intentionally bounded to assignment-related activities within a single course offering. However, despite this domain focus, the extracted logs are not small in an object-centric sense. They capture the complete set of digitally recorded interactions between students, groups, assignments, and teaching staff throughout the course lifecycle, including submissions, grading actions, updates, and group-related interactions.

\begin{table}[t]
	\centering
	\caption{Summary statistics of the extracted OCELs per year}
	\label{tab:ocel_summary}
	\begin{tabular}{lrrrrrr}
		\hline
		Year & Events & Objects & Event types & Object types & E2O & O2O \\
		\hline
		2021 & 60\,797 & 1\,335 & 9 & 6 & 74\,618 & 1\,282 \\
		2022 & 52\,742 & 1\,170 & 8 & 5 & 65\,143 & 1\,150 \\
		2023 & 45\,253 & 1\,309 & 9 & 5 & 56\,630 & 990 \\
		2024 & 38\,313 & 1\,348 & 8 & 5 & 52\,216 & 1\,078 \\
		\hline
	\end{tabular}
\end{table}

The OCELs vary in size and structural properties across years, as summarized in Table~\ref{tab:ocel_summary}. The number of events ranges from 38\,313 to 60\,797 per year, while the number of objects remains relatively stable, between 1\,170 and 1\,348. 
Across all years, the logs contain a stable set of event types (8-9) and object types (5-6), reflecting a consistent domain schema over time. From a structural perspective, the number of event-to-object (E2O) relations is substantial, ranging from 52\,216 to 74\,618 per year, whereas the number of object-to-object (O2O) relations remains comparatively limited (990--1\,282).
This imbalance reflects the event-driven nature of the educational processes, where most complexity stems from events being linked to multiple objects (e.g., students, groups, and assignments).
Consequently, the dataset is well suited for evaluating drill-down and unfold operations that modify event and object type granularity without introducing excessive object-to-object dependency chains.

\subsubsection{Limitations of the Aggregated Process View}\label{sec:explore}
\begin{figure}[h]
	\centering
	\includegraphics[width=\linewidth]{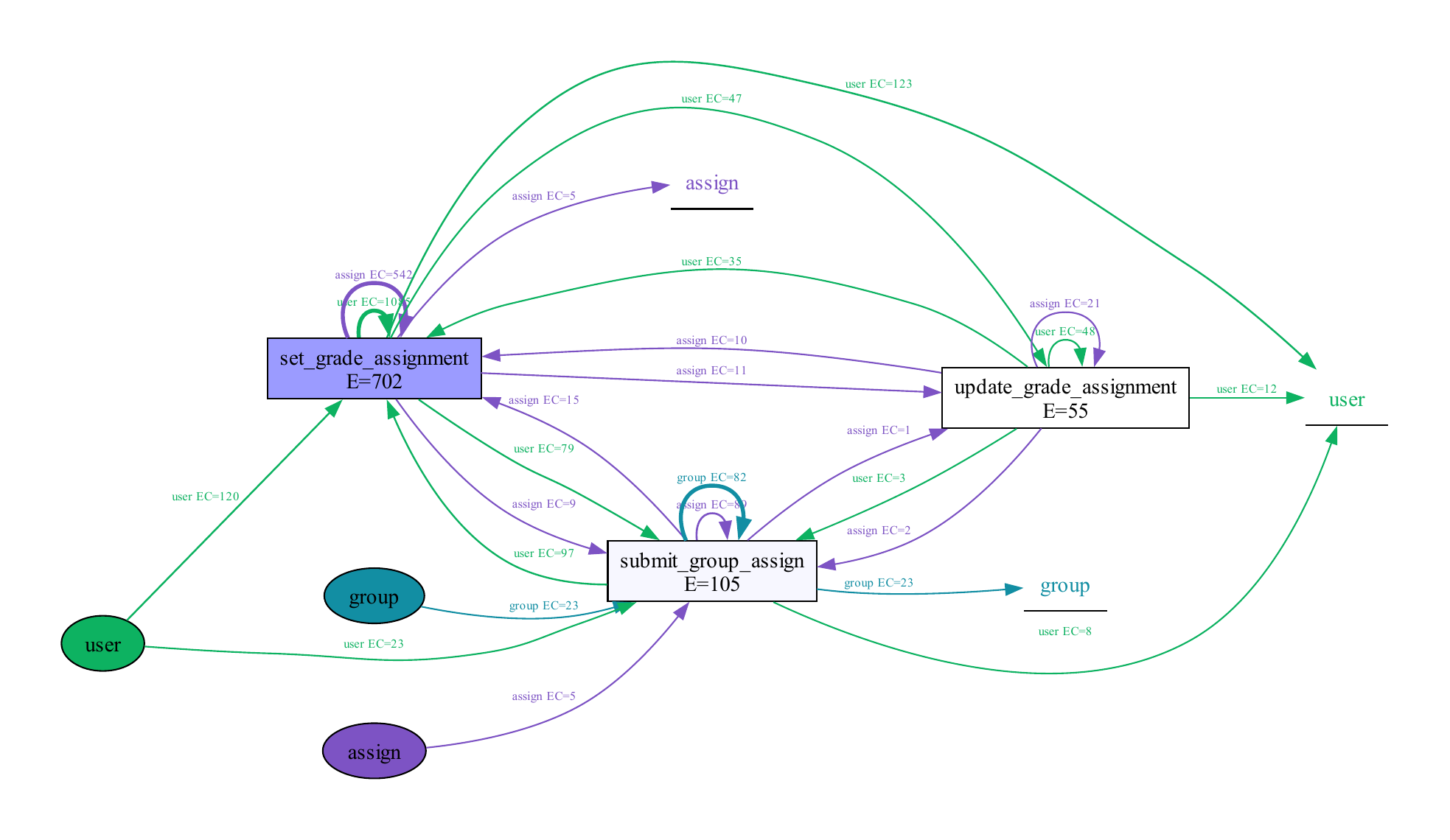}
	\caption{Discovered DFG of 2024 submissions and grading, with limited insights due to coarse granularity}
	\label{fig:overallDFG}
\end{figure}

\figurename~\ref{fig:overallDFG} illustrates the Directly-Follows Graph (DFG), showing how students submitted assignments and how teachers graded them during 2024. As observed, the overall process model is overly abstract and fails to reveal meaningful behavioral patterns. This lack of insight is due to several issues. First, both students and teachers are represented by a single object type, \textit{user}, which obscures their distinct roles. Second, all assignments are grouped under one object type, \textit{assign}, preventing us from distinguishing between specific tasks. Finally, events are aggregated at the event type level, which makes it difficult to identify individual assignment instances. This same pattern was observed in the DFGs for other years, so we do not present them individually here.

\paragraph{Revealing more patterns by changing the log granularity level using OLAP operations}

Following the limitations identified in the previous paragraph, we drilled down into the \textit{user} role attribute and \textit{assign} name to obtain a more detailed view of the process. By distinguishing between students and teachers, we were able to separate their respective activities, thereby addressing the issue of conflating their roles under a single object type. Similarly, drilling down into the \textit{assign} object allowed us to differentiate between individual assignments, which had previously been obscured by the generalized object type. To gain further clarity, we unfolded the events for these drilled-down object types, enabling us to separate events related to each specific user role and assignment instance. This unfolded view revealed a more granular and meaningful process flow, offering deeper insights into the distinct behaviors of students and teachers throughout the assignment submission and grading process.

\figurename~\ref{fig:detailedDFG} illustrates the detailed process flow discovered through our operations. As shown, the process flow includes many steps, making the figure too large to present in full detail in the printed version of this paper without zooming. We have intentionally chosen to display the entire process, highlighting the overall flow consisting of five key steps. Each of these steps represents a milestone that students go through during the course. In the following sections, we will provide a more focused and cropped version of the process for the first and last process steps.

\begin{figure*}[h!]
	\centering
	\includegraphics[width=1\linewidth]{figure_detailed_dfg_2024_steps.pdf}
	\caption{Overall process flow for 2024, intentionally zoomed out to display only the identified key milestones, with details omitted for clarity.}
	\label{fig:detailedDFG}
\end{figure*}

\textbf{Zooming into the First Process Step (Submission and Grading of Process Tree Lab):}\\
\figurename~\ref{fig:detailedDFG_1} presents a zoomed-in view of the first process step cropped from the detailed process flow. This step focuses on the interactions surrounding the Process Tree Lab lab (a sort of assignment), offering a more nuanced understanding of how both students individually and in groups and teachers engage with this specific task.

\begin{figure*}[t!]
	\centering
	\includegraphics[width=1\linewidth]{figure_detailed_dfg_2024_step1.pdf}
	\caption{Zoomed-in view of the 1$^{st}$ process step from ~\figurename~\ref{fig:detailedDFG}}
	\label{fig:detailedDFG_1}
\end{figure*}

The process begins when one student in a group submits the group assignment for the Process Tree Lab, represented by the event \textit{(submit\_group\_assign,(assign\_ProcessTree Lab [pass $|$ fail]))}. This activity is tightly coupled to three distinct object types: the \textit{(user,student)}, the \textit{group} and the \textit{(assign,ProcessTree Lab [pass $|$ fail])}.
As students performed this lab in groups, it was sufficient for one student to submit the lab on behalf of the group, so the relation between \textit{(user, student)} and this event identifies the student who submitted the assignment, and the process in the log starts for these students at this point. 
However, we can see that this is the first event for the whole group. 
By separating these objects, the model allows us to clearly trace which students submitted the assignment.

Once the assignment is submitted, the next phase involves grading. The event \textit{(set\_grade\_assignment, (assign\_ProcessTree Lab [pass $|$ fail]))} captures the teacher's grading activity. This event is connected to both the teacher \textit{(user, teacher)} and the \textit{(user, student)}, showing that the grades are set individually. This is indeed a limitation in Moodle, which registers individual grades for each student even if the assignment is graded for the group. The frequency (E=134) suggests the number of grades set by the teacher in 2024 for students.
We can see that the teacher updated the grades for 19 students later through the \textit{(update\_grade\_assignment, (assign\_ProcessTree Lab [pass $|$ fail]))} event, reflecting the number of students who could not pass the assignment on the first attempt. 
Also, there is a loop on this event for the student object type with frequency of one, showing that one student did not pass the lab on the second attempt and was required to resubmit the lab again.

Interestingly, we can see that the events for resubmission attempts are not captured in the model, indicating that these submissions were handled offline. 
This highlights an important opportunity to improve educational process management by refactoring the process to capture such events.

\textbf{Zooming into the Fifth Process Step (Submission and Grading of Process Mining):}\\
\figurename~\ref{fig:detailedDFG_5} presents a zoomed-in view of the fifth process step cropped from the detailed process flow. This step focuses on the interactions surrounding the Process Mining module. As can be seen, the number of submissions is 13 (much lower compared to 23 Lab submissions). We also see that the assignment was graded for only 28 students, compared to 134 students who received grades for the Process Tree Lab. 
A follow-up investigation showed that the process mining assignment was optional for students; upon completion, they could obtain a better grade for their assignment. 
This shows that the majority of students chose not to participate in this assignment.
Our follow-up shows that this module was scheduled close to the final exam, and many students prioritized preparing for the exam rather than improving their assignment grades.

\begin{figure*}[t!]
	\centering
	\includegraphics[width=1\linewidth]{figure_detailed_dfg_2024_step5.pdf}
	\caption{Zoomed-in view of the 5$^{th}$ process step from ~\figurename~\ref{fig:detailedDFG}}
	\label{fig:detailedDFG_5}
\end{figure*}

In summary, this detailed step uncovers a clearly structured process where students collaborate on submissions, teachers assess and grade the work, and grades are subsequently updated. It demonstrates how OLAP operations like drilling down and unfolding not only improve visibility but also make the coordination between different actors and objects explicit, facilitating meaningful process analysis.

\subsubsection{Precision and fitness evaluation}\label{sec:precfit}

To evaluate the impact of our approach on the precision and fitness of discovered models, we used the \texttt{ocpa} library~\cite{adams2022ocpa}, capable of discovering object-centric Petri nets~(OCPN) and calculating their fitness and precision. However, the library currently has two limitations: (i) it cannot calculate the fitness and precision of large logs or complex processes, and (ii) it only supports OCEL 1.0. To mitigate these limitations, we (i) extracted separate OCEL files for each group, capturing the students' work within their groups, and (ii) converted the logs to OCEL 1.0 format.

Using these logs, we discovered object-centric Petri nets and calculated the fitness and precision for each group. Subsequently, we drilled down the logs to (i) separate teachers from students, as both were recorded under the same user object type, and (ii) distinguish submission boxes for different assignments, which were expected at various stages of the course. Finally, we unfolded the logs for these submission boxes. The transformed logs were then used to discover new object-centric Petri nets and recalculate their fitness and precision.

\begin{figure}[h]
	\centering  
	\includegraphics[width=0.8\textwidth]{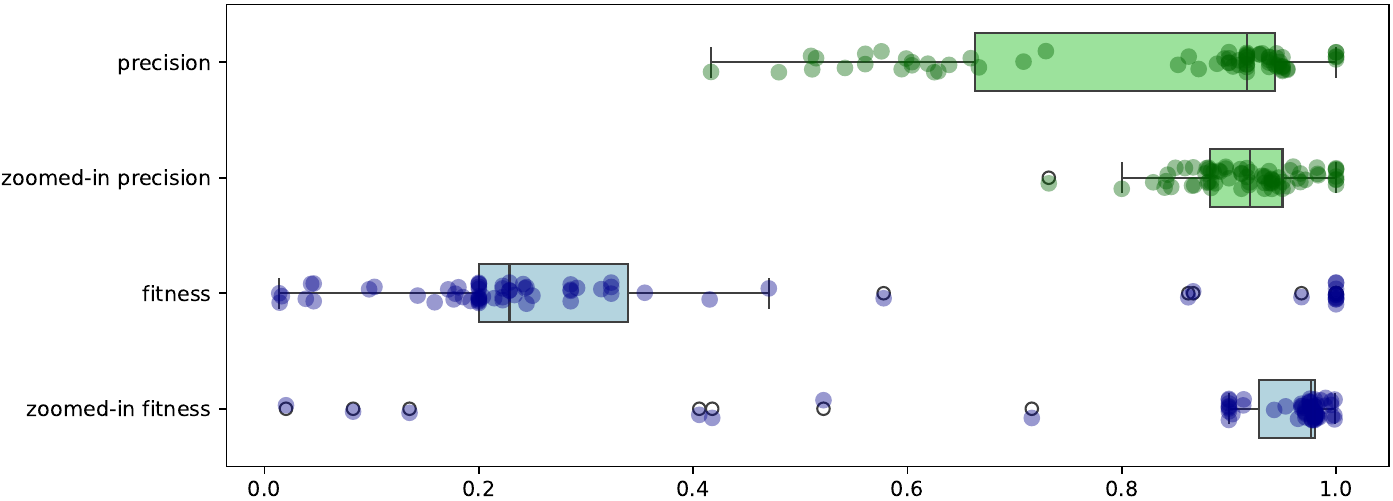}
	\caption{Comparison of precision and fitness for discovered object-centric Petri nets based on original vs. drilled-down and unfolded logs.}
	\label{fig:fitness}
\end{figure}

\figurename~\ref{fig:fitness} shows the precision and fitness of the extracted logs with respect to discovered models compared to the transformed versions for 71 groups. 
The \texttt{ocpa} library timed out when computing fitness and precision for 20 out of 91 groups due to model/log complexity. Consequently, the quantitative comparison in \figurename~\ref{fig:overall_case} is based on the remaining 71 groups. The OLAP operations themselves were applied to all groups; only the conformance computation failed on this subset. We therefore interpret the reported fitness/precision improvements as conservative, restricted to the subset of groups for which conformance checking is feasible with current tooling.
As can be seen, the fitness and precision improved significantly for most groups after drilling down and unfolding the logs. However, five groups exhibited low fitness scores (below 0.5).

\subsubsection{Outlier and error analysis}\label{sec:erroranalysis}

\paragraph{Outlier analysis:}
To investigate these low scores, we analyzed the tEKG created from the original OCEL 2.0 logs~\cite{khayatbashi2024transforming} after applying the OLAP operations described in Section~\ref{sec:Approach}. Representing the data as a temporal event knowledge graph allowed us to follow student-group relations as explicit, time-stamped edges and to query and visualize how these relations evolved over time.

The two groups with the lowest fitness scores participated in the course in 2024. These groups exhibited a high degree of rolling membership, where students frequently switched groups during the course. Such changes likely contributed to misalignments, as current OCPM techniques do not account for dynamic object-to-object relationships when discovering process models. Consequently, the entire process was discovered based on the final group to which each student belonged rather than the active group at the time of specific events.

\begin{figure}[t!]
	\centering  
	\includegraphics[width=1\textwidth]{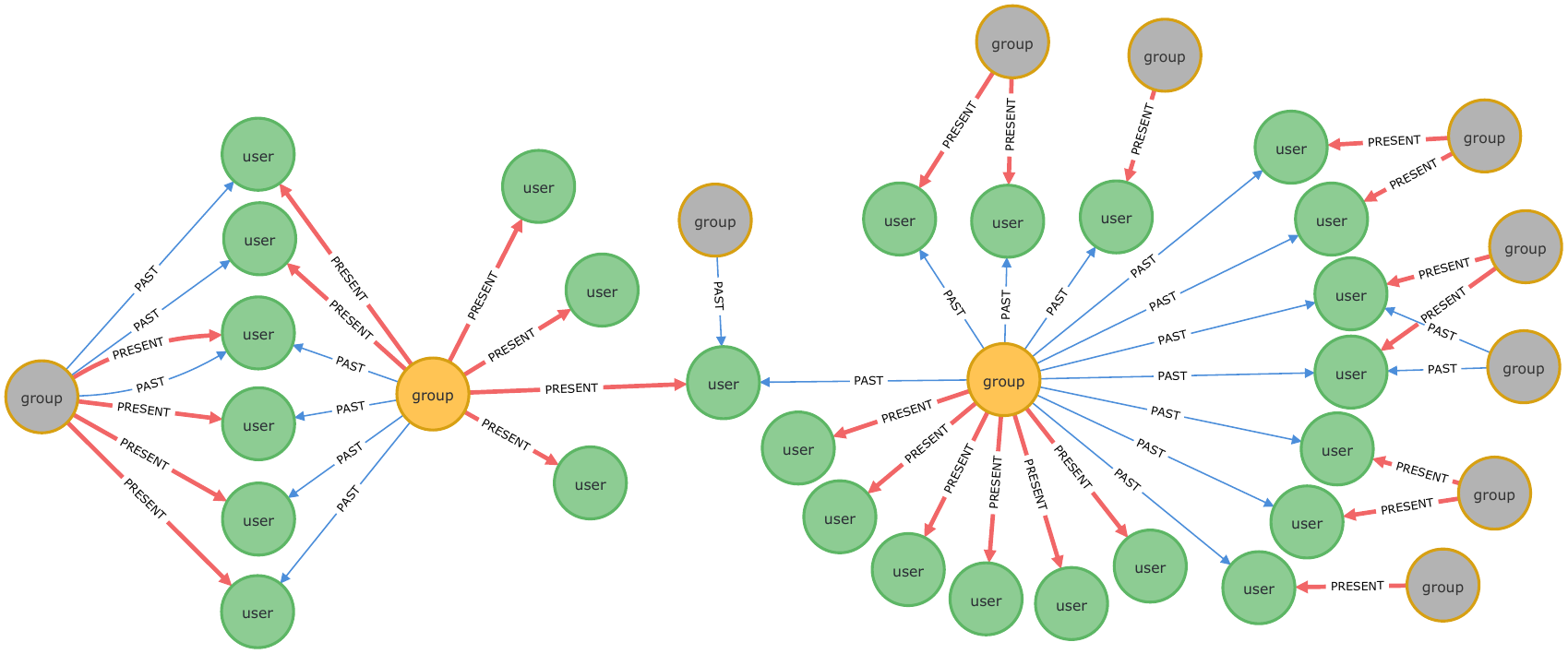}
	\caption{The relation between students and two groups with the lowest fitness score showing the high group dynamics within these groups.}
	\label{fig:neo4j}
\end{figure}

To better illustrate this issue, we modified the graph to differentiate between students' current and former groups. In the tEKG, these relationships were labeled \texttt{REL}; for demonstration purposes, we relabeled them as \texttt{PRESENT} and \texttt{PAST} and manually applied distinct colors.
\figurename~\ref{fig:neo4j} visualizes the rolling membership for the two groups with the lowest fitness scores, highlighted as yellow nodes. The yellow group on the right has the lowest fitness score, with 11 students leaving to join other groups, while the yellow group on the left saw four departures, which is still substantial given its size of six members. These significant changes in group composition likely explain the observed low fitness scores.

The transformation of OCELs into a tEKG demonstrates the importance of converting object-centric event data between formats~\cite{khayatbashi2023transforming,khayatbashi2024transforming}. In particular, the graph representation enables analyses that are difficult to perform on the OCEL alone, such as tracing temporal paths of students across multiple groups or computing graph-based measures of group dynamics. Such transformations enable analysts to leverage the strengths of various tools provided by different data formats, thereby enhancing process analysis capabilities.

\paragraph{Error analysis:}
We analyzed the groups for which errors occurred during the calculation of fitness and precision. We found that 10 out of the 20 problematic groups were from 2023. To investigate this issue further, we discovered OC-DFGs for each year, both before and after log transformation.\footnote{Case study materials including OC-DFGs are available at \url{https://github.com/shahrzadkhayatbashi/olap-operations4ocel/blob/main/experiment-results/case_study_summary.md}} Our analysis revealed that the process model for 2023 was significantly more complex compared to other years. This insight could not be identified without drilling down and unfolding the logs.

The root cause of this complexity was a change introduced to the course in 2023, where optional tracks were implemented. This adjustment led to a less structured process, as students followed different tracks, increasing the complexity of the workflow. In 2024, this change was revoked, restoring a more structured and uniform learning path. This experience highlights the need for more advanced OCPM algorithms capable of handling knowledge-intensive and less-structured processes effectively.

To better understand how structural characteristics relate to discovery errors, we computed point-biserial correlations between the error variable and several OCPN complexity metrics such as number of places, transitions, arcs, silent transitions, `AND' split and joins, and `XOR' split and simple merge in discovered models. All these attributes showed moderate positive correlations with error: places ($r = 0.527$, $p < 0.001$), transitions ($r = 0.505$, $p < 0.001$), arcs ($r = 0.523$, $p < 0.001$), silent transitions ($r = 0.487$, $p < 0.001$), AND-splits ($r = 0.490$, $p < 0.001$), AND-joins ($r = 0.484$, $p < 0.001$), XOR-splits ($r = 0.475$, $p < 0.001$), and simple merges ($r = 0.465$, $p < 0.001$). 
These results confirm that higher structural complexity of the discovered OCPNs is systematically associated with a higher probability of conformance checking errors.

\begin{figure}[t]
	\centering
	\includegraphics[width=0.65\linewidth]{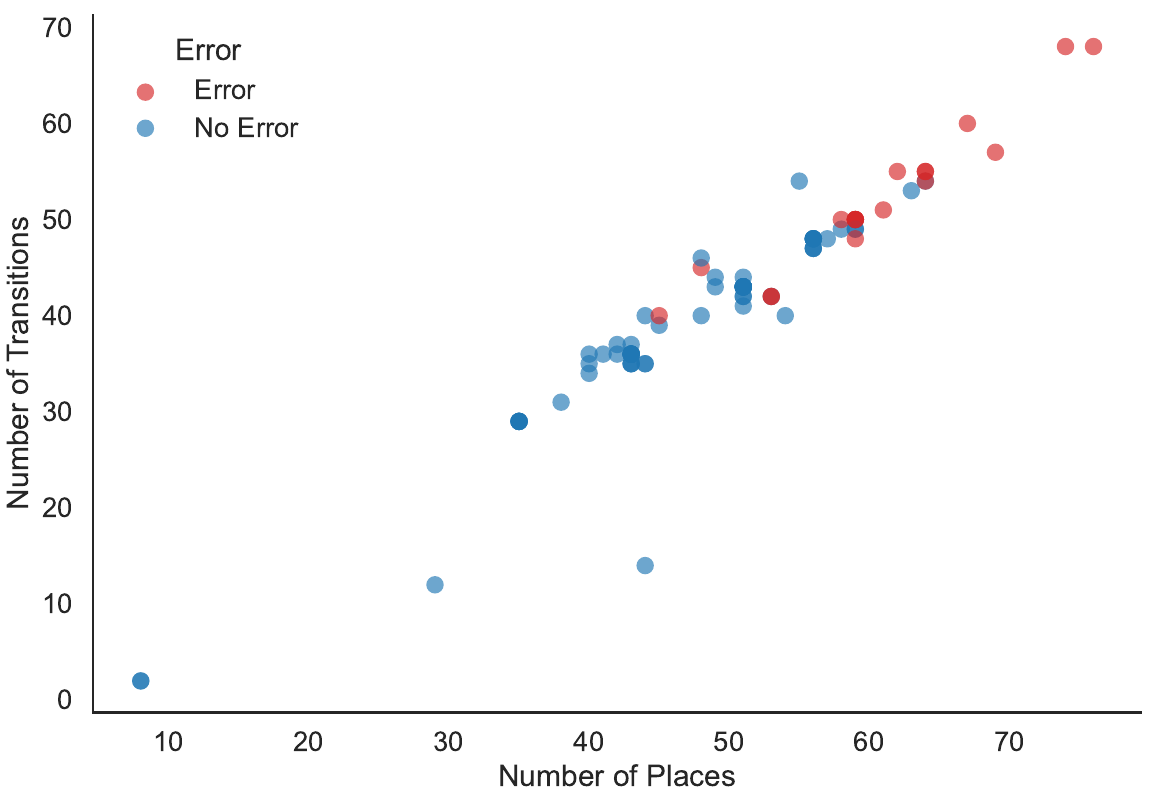}
	\caption{Relationship between the number of places and transitions in the discovered OCPNs, with errors occurring when computing fitness and precision in larger models.}
	\label{fig:place_transitions}
\end{figure}

\figurename~\ref{fig:place_transitions} visualizes the relationship between the number of places and the number of transitions in the discovered OCPNs. The plot shows a clear positive association: models with more places also tend to contain more transitions, reflecting the overall growth of structural complexity. Error cases are more frequently observed among models with higher values on both axes, indicating that larger and structurally richer nets are more prone to discovery failures. Because many points overlap in the upper region of the plot, making it difficult to visually disentangle the contribution of specific constructs, we further examined how errors relate to the number of AND-splits and AND-joins in the models.

\begin{figure}[t]
	\centering
	\includegraphics[width=0.65\linewidth]{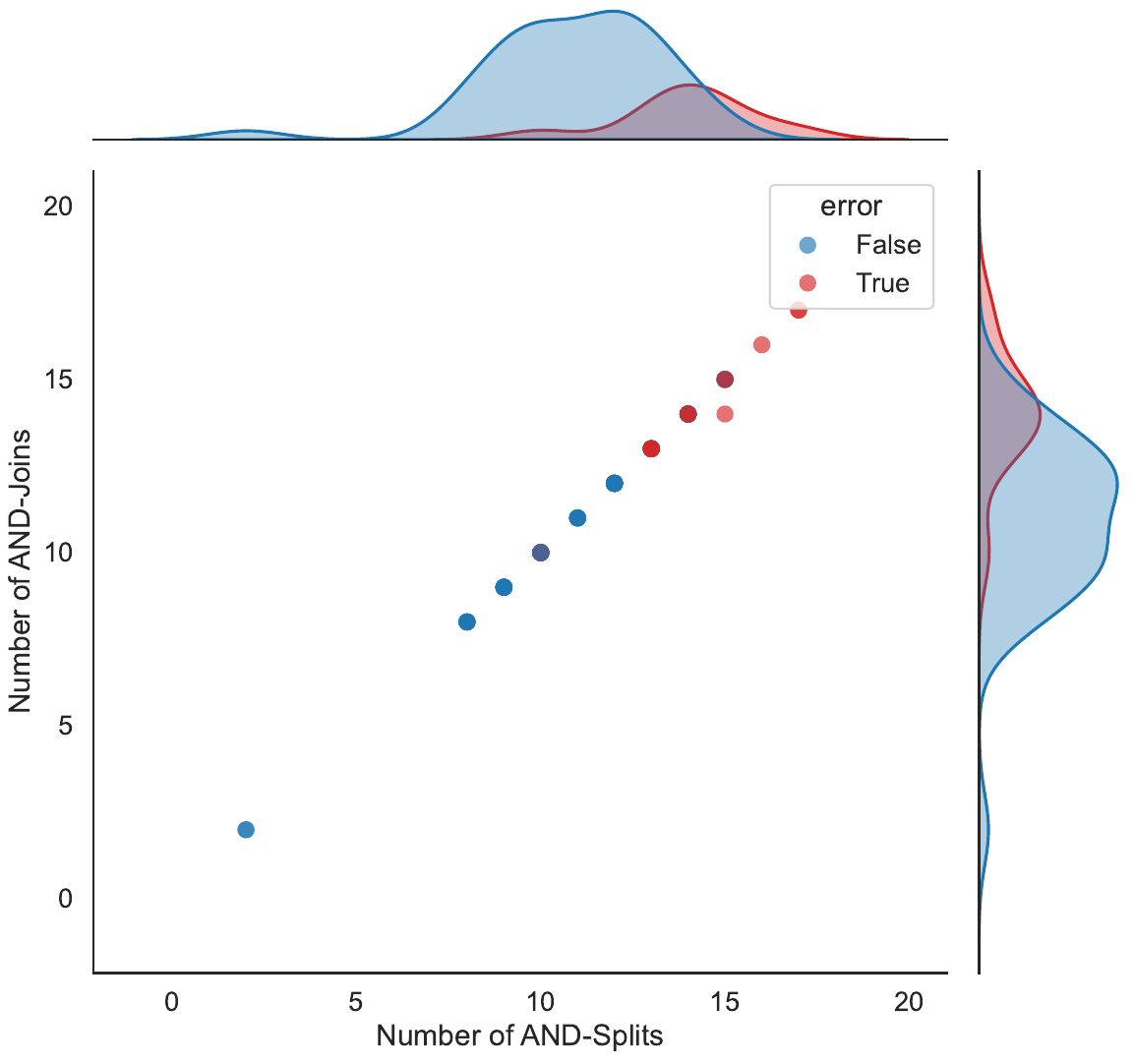}
	\caption{Relationship between the number of parallel splits and joins in the discovered OCPNs, with errors occurring when computing fitness and precision in larger models.}
	\label{fig:and_splits_joins}
\end{figure}

Figure~\ref{fig:and_splits_joins} provides a more detailed view of synchronization-heavy models by plotting the number of parallel splits and joins in the discovered models. The processes for which the fitness and precision could not be computed tend to have higher levels of parallelism. A plausible explanation is that some student groups did not work collaboratively, but instead divided the work and executed activities in parallel. This also suggests that models with substantial synchronization behavior are particularly prone to failures in calculating fitness and precision, indicating a need for more robust and scalable conformance-checking algorithms.

\begin{figure}[t]
	\centering
	\includegraphics[width=0.65\linewidth]{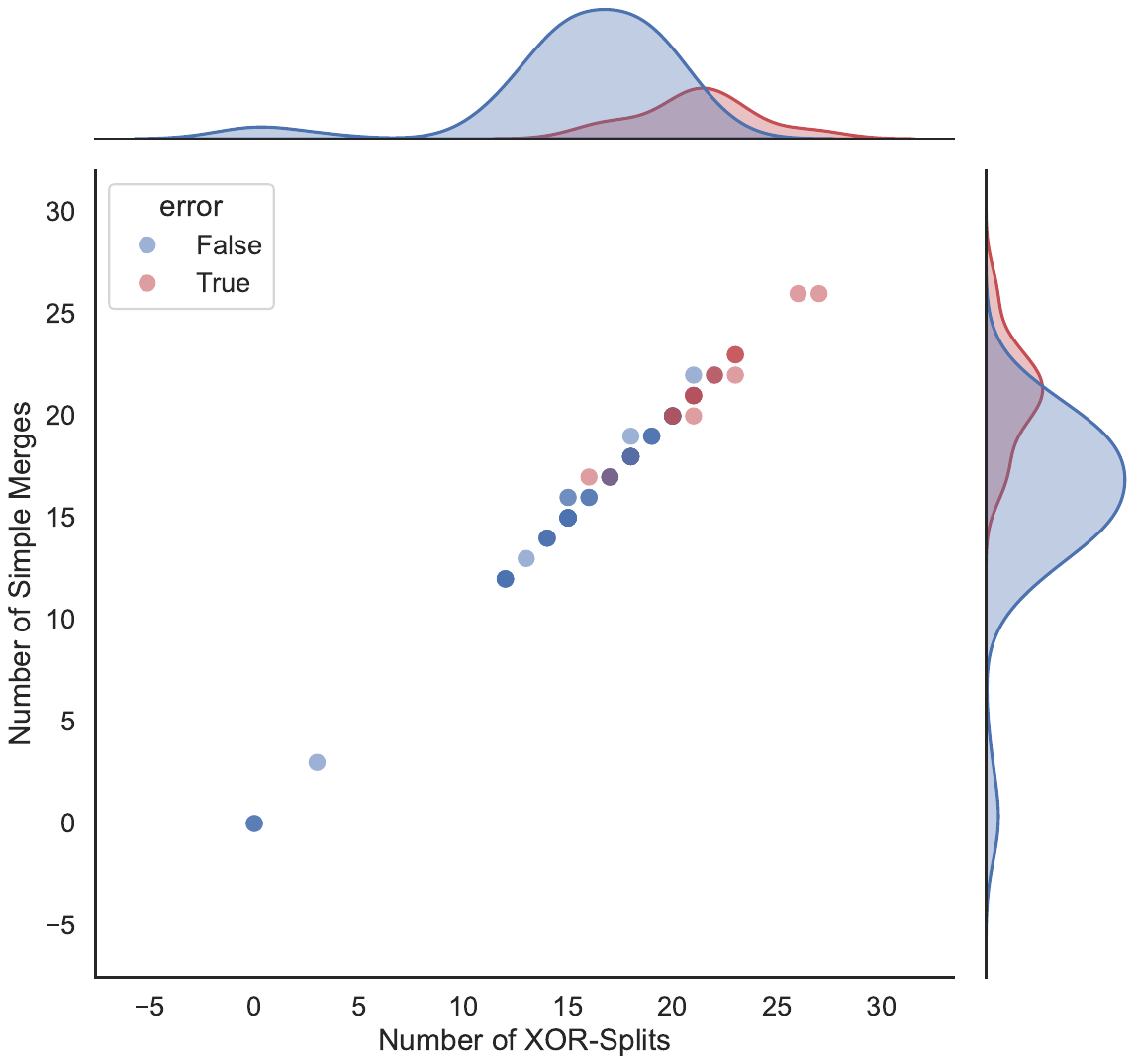}
	\caption{Relationship between the number of XOR and simple merge in the discovered OCPNs, with errors occurring when computing fitness and precision in larger models.}
	\label{fig:xor_splits_merges}
\end{figure}

In contrast, the relationship between XOR-splits and simple merges, shown in Figure~\ref{fig:xor_splits_merges}, is noticeably weaker. Although error cases still tend to appear in models with more branching and merging at places, the overlap between error and non-error cases is larger, and the separation in the marginal distributions is less pronounced. This suggests that XOR-based choice structures play a secondary role in error occurrence and are not as strongly associated with failures as synchronization patterns.

Overall, these results demonstrate that structural complexity (arising from model size, branching, synchronization, and merging) collectively contributes to errors in calculating fitness and precision. However, synchronization-heavy constructs (AND-splits and AND-joins) appear to be the most influential, while XOR-based constructs exert a more modest effect.

\subsubsection{Performance Evaluation in the Educational Case Study}

The OCELs in our case study are comparable in size to the per-municipality OCELs in BPIC~15~\cite{ocel15}. To evaluate the performance of the proposed granularity-changing operations in this context, we applied a controlled drill-down and unfold operation to all yearly OCELs used in the case study. Specifically, we performed a drill-down on the \texttt{assign} object type using the assignment identifier as the drilling attribute, thereby distinguishing individual assignments at the object type level. Subsequently, we unfolded all event types that were related to assignments, allowing event semantics to be specialized with respect to the drilled assignment instances. This configuration reflects a realistic analytical scenario in which an analyst seeks to explore assignment-specific behavior while preserving the overall process context.

In the context of this evaluation, we refer to each combination of a drilled object type and an unfolded event type as an \emph{unfold call}. Each unfold call corresponds to a single invocation of the unfold operation that materializes event semantics for a specific event-object type combination. In practical analysis settings, analysts rarely apply drill-down and unfold operations exhaustively across all possible combinations. Instead, they typically zoom in on selected object types or event types that are of immediate analytical interest. In this experiment, however, we deliberately applied unfold operations across all eligible combinations in order to assess the performance characteristics of the approach under increased granularity demands and to establish a conservative upper bound on runtime behavior.

Table~\ref{tab:granularity_expansion} summarizes the structural impact of applying drill-down and unfold operations across all eligible combinations. Across all years, exactly one object type was drilled down, corresponding to the assignment dimension. The number of object types involved in unfold operations ranges from 9 to 12, depending on the year, reflecting differences in the number of distinct assignments submitted by students. The number of event types considered for unfolding remains stable between 8 and 9. As a result of these operations, the total number of event types increases substantially, reaching between 50 and 72, while the number of object types increases to between 13 and 16.

\begin{table}[h]
	\centering
	\caption{Granularity expansion induced by drill-down and unfold operations}
	\label{tab:granularity_expansion}
	\begin{tabular}{lrrrrrrrr}
			\hline
			Year &
			\multicolumn{3}{c}{Number of} &
			Drilled &
			Unfold &
			Unfold &
			After &
			After \\
			& 
			Events &
			Objects &
			E2O &
			OT &
			OT &
			ET &
			ET &
			OT \\
			\hline
			2021 & 60\,797 & 1\,335 & 74\,618 & 1 &  9 & 9 & 69 & 14 \\
			2022 & 52\,742 & 1\,170 & 65\,143 & 1 & 11 & 8 & 72 & 15 \\
			2023 & 45\,253 & 1\,309 & 56\,630 & 1 & 12 & 9 & 72 & 16 \\
			2024 & 38\,313 & 1\,348 & 52\,216 & 1 &  9 & 8 & 50 & 13 \\
			\hline
		\end{tabular}
\end{table}

These results illustrate that even a single drill-down operation can induce a considerable expansion of the event and object type space when combined with unfold operations. Moreover, the observed variation across years demonstrates that the degree of granularity expansion is influenced not only by the number of base types but also by their distribution and associations within the log. Importantly, the number of types after expansion remains manageable, indicating that additional drill-down steps are still feasible when applied selectively.

The runtime implications of this granularity expansion are reported in Table~\ref{tab:runtime_summary}. The table decomposes the total processing time into log reading, drill-down, and unfold phases. Log reading (performed using PM4Py) requires between 2.512 and 6.429 seconds per year, reflecting differences in file size and serialization overhead. Across all years, drill-down remains lightweight, with execution times between 0.109 and 0.295 seconds, which is consistent with the fact that it only refines object types based on recorded attribute values.

\begin{table}[h]
	\centering
	\caption{Runtime characteristics of reading, drill-down, and unfold operations (per year)}
	\label{tab:runtime_summary}
	\begin{tabular}{lrrrrrr}
		\hline
		Year &
		Read &
		Drill-down &
		Unfold &
		Unfold &
		Avg unfold &
		Avg unfold \\
		&
		(s) &
		(s) &
		(s) &
		calls &
		per call (s) &
		per ET (s) \\
		\hline
		2021 & 6.094 & 0.138 & 5.955 & 81  & 0.074 & 0.662 \\
		2022 & 6.429 & 0.295 & 13.154 & 88 & 0.149 & 1.644 \\
		2023 & 2.512 & 0.109 & 8.275 & 108 & 0.077 & 0.919 \\
		2024 & 3.766 & 0.113 & 4.039 & 72  & 0.056 & 0.505 \\
		\hline
	\end{tabular}
\end{table}

Total unfold time ranges from 4.039 to 13.154 seconds, driven by the number of unfold calls required to materialize event semantics for all event--object type combinations. The number of unfold calls varies between 72 and 108 per year, closely mirroring the degree of type expansion reported in Table~\ref{tab:granularity_expansion}. Despite this variation, the average runtime per unfold call remains stable and sub-second, ranging from 0.056 to 0.149 seconds. When aggregated per event type, unfold costs range from 0.505 to 1.644 seconds.

\begin{figure}[t]
	\centering
	\includegraphics[width=\linewidth]{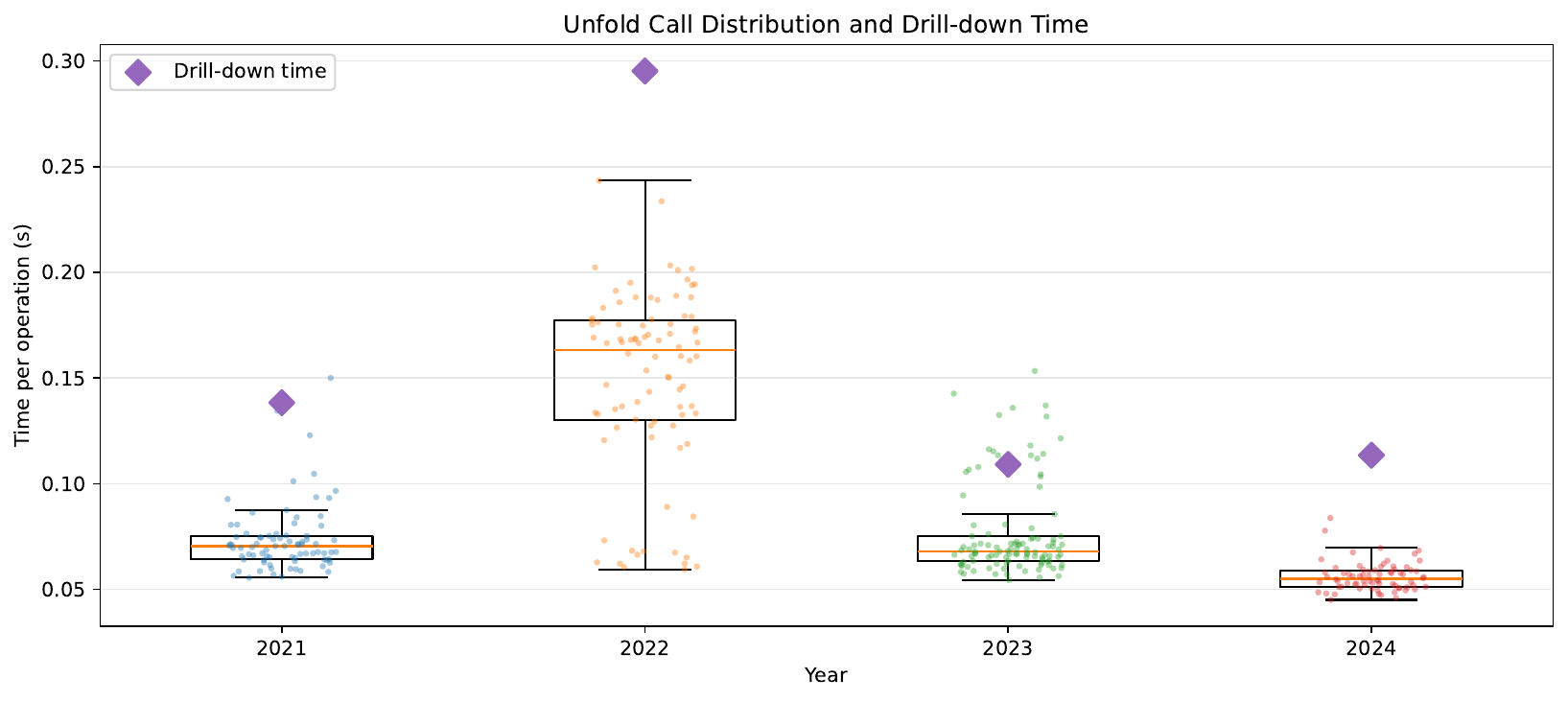}
	\caption{Distribution of unfold runtimes per call across years for the case study logs. Boxplots summarize per-call unfold time, points show individual unfold calls, and diamond markers indicate drill-down time.}
	\label{fig:unfold_call_dist}
\end{figure}

\figurename~\ref{fig:unfold_call_dist} provides a finer-grained view of the runtime behavior by reporting the distribution of unfold runtimes at the level of individual unfold calls. For each year, the boxplot summarizes the variability of per-call unfold time, while the overlaid points show the raw runtimes of all unfold calls. The diamond markers indicate the corresponding drill-down time. Two observations follow. First, unfold calls exhibit stable sub-second runtimes across all years, with moderate dispersion and only a limited number of slower calls. Second, drill-down remains consistently small compared to the cumulative unfold phase and, importantly, compared to the OCEL loading time reported in Table~\ref{tab:runtime_summary}.

Overall, these results show that the computational overhead of applying drill-down and unfold operations is modest relative to the cost of reading and parsing OCEL logs using PM4Py. This is relevant for practical deployments, where OCEL loading constitutes an unavoidable baseline cost before any analytical transformation can be executed. Moreover, because we intentionally executed unfold calls for all eligible event--object type combinations, the reported distributions reflect a conservative upper bound on runtime behavior under increased granularity demands. In typical interactive analysis, where an analyst selects only a subset of event types or object types for zooming, the expected runtime would be correspondingly lower.

\subsection{Scalability Evaluation on Public Datasets}
\label{sec:performanceeval}

While the case study logs capture realistic usage scenarios, their size may not be representative of large industrial deployments. To complement the case study and to support reproducibility, we conducted a second experiment on publicly available OCELs derived from the BPIC~2014~\cite{ocel14}, BPIC~2016~\cite{ocel16}, BPIC~2017~\cite{ocel17}, and BPIC~2019~\cite{ocel19} collections. BPIC~2015~\cite{ocel15} was omitted, as its log is not sufficiently large for a meaningful performance evaluation. All these OCELs were transformed from Event Knowledge Graphs and are publicly available, which makes the experiment reproducible~\cite{khayatbashi2023transforming}~\footnote{The evaluation code is available at \url{https://github.com/shahrzadkhayatbashi/olap-operations4ocel}}.

For this experiment, we applied the same conceptual configuration as in the case study: a drill-down on the most frequent object type, followed by unfold operations for all event types strongly associated with that object type. In the original Event Knowledge Graphs, the BPIC objects did not have any attributes to be used for drill-down. Therefore, in the corresponding BPIC OCELs we introduced a synthetic attribute on the selected object type with $n$ distinct values, distributed proportionally across the objects. To systematically assess the effect of the number of categories on performance, we varied $n$ from 2 to 10 and measured drill-down time and unfold time per call. For each BPIC year and each value of $n$, we averaged these measurements over all OCEL files of that year.

\begin{figure}[t]
	\centering
	\begin{subfigure}[t]{0.48\linewidth}
		\centering
		\includegraphics[width=\linewidth]{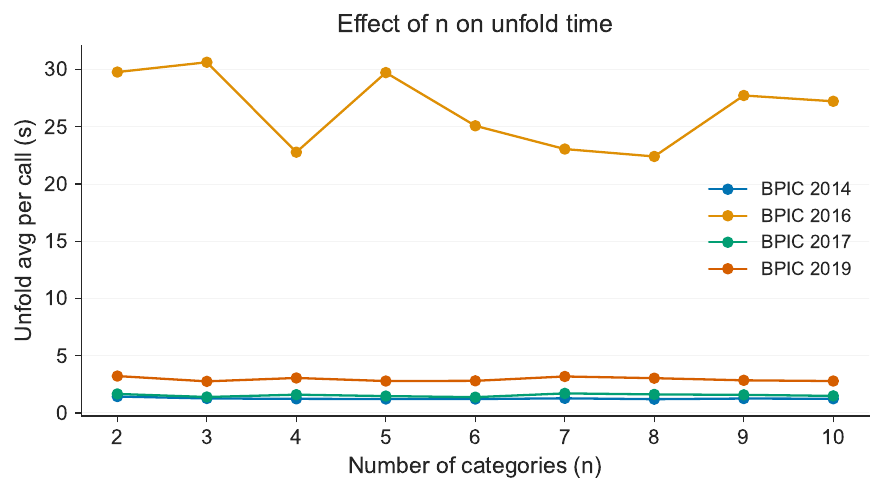}
		\caption{Average unfold time per call as a function of the number of categories ($n$) for the four BPIC logs.}
		\label{fig:bpic-effect-n-unfold}
	\end{subfigure}
	\hfill
	\begin{subfigure}[t]{0.48\linewidth}
		\centering
		\includegraphics[width=\linewidth]{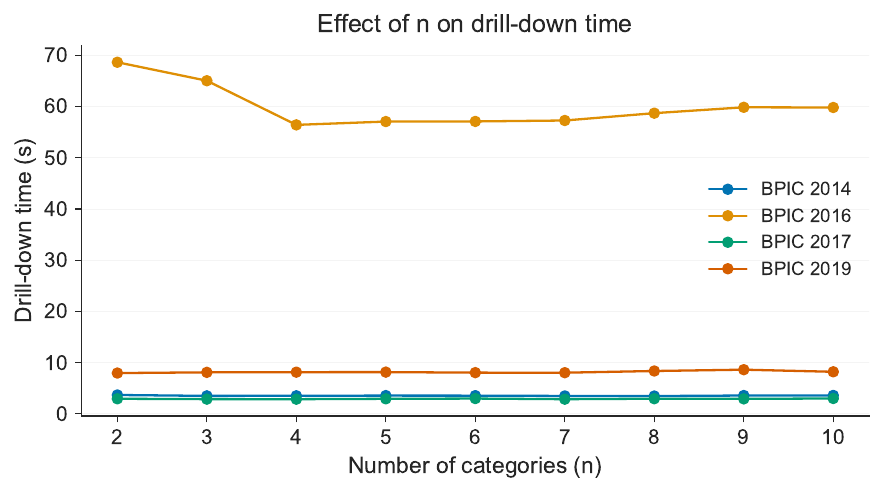}
		\caption{Average drill-down time as a function of the number of categories ($n$) for the four BPIC logs.}
		\label{fig:bpic-effect-n-drilldown}
	\end{subfigure}
	\caption{Effect of the number of categories ($n$) on unfold and drill-down time for the BPIC logs.}
	\label{fig:bpic-effect-n}
\end{figure}

Figures~\ref{fig:bpic-effect-n-unfold} and~\ref{fig:bpic-effect-n-drilldown} show the average unfold time per call and drill-down time, respectively, as a function of~$n$. Across all four BPIC logs, the curves do not show any systematic upward trend, in either metric. This suggests that, in our setting, the number of categories has a limited influence on performance compared to other factors such as number of objects and events.

Given this observation, we fix \(n = 10\) for the remainder of the analysis and focus on how performance depends on the characteristics of each log. Table~\ref{tab:bpic-summary} summarizes, for each BPIC year, the number of events, objects, and event--object relations together with the \emph{average} read time, unfold time per call, and drill-down time at \(n = 10\) (averaged over all OCEL files per year). The four logs range from roughly seven hundred thousands to more than seven million events and from roughly one hundred thousand to more than seven hundred thousand objects. As expected, the largest log, BPIC~2016, incurs the highest unfold and drill-down times, and it also exhibits by far the longest read time when importing the OCEL into PM4Py. 

\begin{table}[h]
	\centering
	\caption{Log sizes and average performance metrics for the BPIC logs. Values for time are averaged over all OCEL files per year.}
	\label{tab:bpic-summary}
	\begin{tabular}{lrrrrrr}
		\toprule
		BPIC & Events & Objects & Relations & Read (s) & Unfold (s) & Drill-down (s) \\
		\midrule
		BPIC 2014 &   690\,622 & 228\,885 &  2\,732\,213 &  34.7 &  1.43 &  3.55 \\
		BPIC 2016 & 7\,360\,146 & 748\,913 & 36\,430\,880 & 659.9 & 29.78 & 59.97 \\
		BPIC 2017 & 1\,202\,267 & 106\,162 &  2\,404\,534 &  48.8 &  1.65 &  2.93 \\
		BPIC 2019 & 1\,595\,923 & 330\,685 &  5\,984\,602 &  80.2 &  3.22 &  8.18 \\
	\end{tabular}
\end{table}

\begin{figure}[t]
	\centering
	\includegraphics[width=0.9\linewidth]{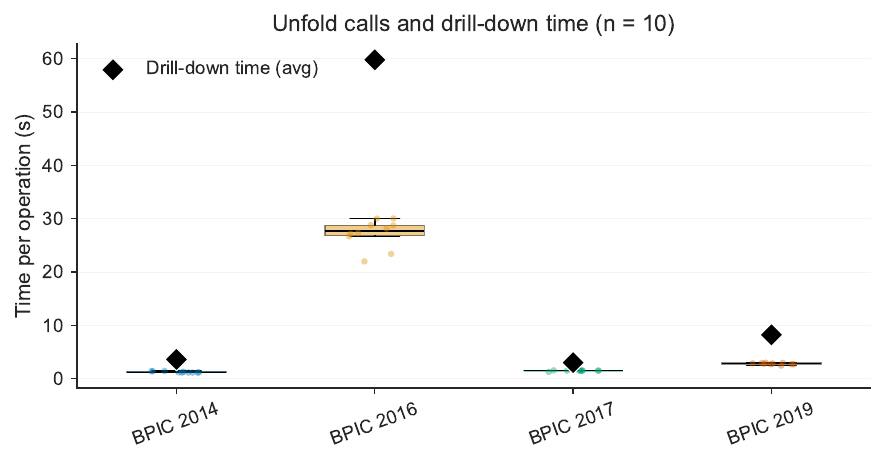}
	\caption{Distribution of unfold runtimes per call and average drill-down times for BPIC~2014/2016/2017/2019 (\(n = 10\)). }
	\label{fig:bpic-unfold-distribution}
\end{figure}

To obtain a more fine-grained view of runtime variability on the BPIC logs, we also analyze the distribution of individual unfold calls for \(n = 10\).
Figure~\ref{fig:bpic-unfold-distribution} reports the distribution of unfold runtimes at the level of individual unfold calls for each BPIC log. For each log, the boxplot captures the spread of per-call unfold times, while the overlaid points show all raw call durations. The diamond markers indicate the corresponding per-year average drill-down time.

Two observations follow. First, for BPIC~2014, BPIC~2017, and BPIC~2019, unfold calls remain tightly clustered around a characteristic value for each log: roughly \(1.5\)~seconds per call for BPIC~2014/2017 and around \(3.2\)~seconds for BPIC~2019, with only moderate dispersion. This aligns with Table~\ref{tab:bpic-summary}, where these three logs differ in size by at most a factor of about 2-3. 
Second, BPIC~2016 stands out as a genuine stress test: its unfold calls concentrate around \(29\)~seconds, and drill-down time is around \(60\)~seconds on average, which is consistent with its much larger number of events and, in particular, its very high number of event--object relations.
Importantly, however, Table~\ref{tab:bpic-summary} shows that the read time for BPIC~2016 is an order of magnitude larger than for the other logs. Hence, most of the absolute runtime overhead for BPIC~2016 is already incurred when loading the OCEL via PM4Py. 
This finding also underscores the importance of supporting roll-up and fold operations, allowing analysts to reverse zoom-in actions without reloading the log. Reloading is roughly ten times slower than drill-down and even more costly than unfolding, making in-place abstraction operations crucial for interactive analysis.

\begin{figure}[h]
	\centering
	\includegraphics[width=0.7\linewidth]{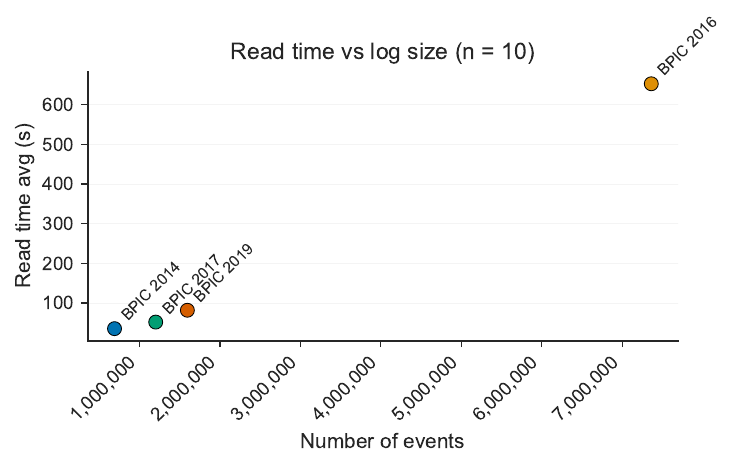}
	\caption{The read time using PM4Py vs.\ log size (number of events) for the BPIC logs.}
	\label{fig:bpic-read-vs-events}
\end{figure}

Finally, we relate performance directly to log size. To better disentangle the different contributions to the overall runtime, we first relate the average read time to log size.
Figure~\ref{fig:bpic-read-vs-events} plots the average time required to load the OCEL into PM4Py against the number of events for \(n = 10\).
BPIC~2016 clearly stands out: its read time is roughly an order of magnitude higher than for the other logs, mirroring its substantially larger number of events and relations.
This indicates that a considerable portion of the absolute runtime is already spent in the underlying OCEL loading and parsing, before any drill-down or unfold operation is executed.

Overall, these plots confirm that:
(i) for very large logs such as BPIC~2016, the dominant cost is reading the OCEL into PM4Py, 
(i) performance scales primarily with log size and connectivity, and
(iii) the additional overhead introduced by the proposed drill-down and unfold operations remains moderate.
Together, this supports the suitability of the granularity-changing operations for interactive analysis on both moderate and large real-life OCELs.

\subsection{Threats to validity}
Our evaluation of fitness and precision is subject to a technical limitation. For 20 out of 91 groups, the \texttt{ocpa} library timed out when calculating fitness and precision due to the size and complexity of the corresponding logs and models (see Section~\ref{sec:precfit}). As a result, for these groups we could not quantify how fitness and precision changed before and after applying the proposed operations. This limitation affects only the completeness of the quantitative conformance analysis. The OLAP operations are defined at the log level and were successfully applied to all 91 groups. 

Likewise, the exploratory analyses and process discovery results reported in Sections~\ref{sec:explore}--\ref{sec:performanceeval} include all groups and all years. In other words, the timeout in the conformance checking library restricted our ability to measure fitness and precision improvements for a subset of groups but did not limit the applicability of the proposed approach for process exploration. More broadly, this observation underlines current scalability constraints of object-centric conformance checking. Developing more scalable and robust methods to calculate fitness and precision for object-centric Petri nets remains an important direction for future research.

The evaluation is based on data from a single course in one university, which may limit the generalizability of the empirical findings. Although the underlying OCELs cover four academic years and 91 groups and the same operations have been applied in other domains in previous work, the quantitative results reported here are context-specific and may not directly transfer to other processes, domains, or logging conventions. Additional case studies and benchmarks on heterogeneous datasets will be needed to more systematically assess the external validity of the approach.

Finally, the core of our approach is to alter the event log, not the discovery algorithm. Consequently, the precision and fitness values we report compare models discovered from different versions of the log (original vs. drilled-down and unfolded). These metrics therefore do not evaluate the intrinsic quality of a fixed discovery algorithm on a fixed log, but rather the effect of changing the granularity and structure of the input data on the resulting models. While this is in line with our goal of studying the impact of granularity-changing operations, it should be kept in mind when interpreting the conformance results: the improvements observed reflect both the ability of the discovery algorithm to exploit the refined log and the analyst's choice of granularity configuration.


\section{Conclusion}\label{sec:Conclusion}

This paper introduced and demonstrated the application of drill-down, roll-up, unfold, and fold operations in Object-Centric Process Mining (OCPM). By implementing these operations within the OCEL 2.0 standard, we enable precise and multi-dimensional analysis of business processes. Our approach was validated through a real-world case study, where drill-down and unfold operations significantly improved the fitness and precision of discovered models, underscoring the practical applicability and effectiveness of our approach.

We further demonstrated the transformation of OCELs into temporal Event Knowledge Graphs (tEKG), illustrating how these complementary data representations can enhance process analysis. In our case study, groups with high rolling membership challenged existing OCPM techniques, as they struggled to capture dynamic object-to-object relationships. By combining OCEL and tEKG, we revealed the potential for deeper analysis and richer insights in such scenarios.
The results emphasize that the proposed OLAP operations provide the flexibility needed for comprehensive process analysis, enabling analysts to uncover intricate patterns and variations often missed by traditional single-case process mining techniques. These findings highlight the importance of integrating multi-dimensional operations into OCPM to advance the scope and precision of process insights.

Future research should address several key directions, including: (i) developing scalable methods for calculating fitness and precision for object-centric Petri nets, and (ii) improving techniques for capturing dynamic object-to-object relationships during process model discovery. Additionally, integrating these operations into existing process mining tools could enhance their functionality, providing robust support for analyzing complex and dynamic processes.
As OCPM evolves, the incorporation of drill-down and roll-up operations will be pivotal in advancing the depth and quality of insights derived from multi-dimensional process data.

In conclusion, this research establishes a foundation for more detailed and dynamic process mining methodologies, opening new pathways for organizational efficiency and innovation through data-driven process analysis.

\backmatter

\section*{Declarations}

\begin{itemize}
\item Funding: Not applicable
\item Conflict of interest/Competing interests: The authors declare no Conflict of interest.
\item Data availability: The test data is available at \url{https://github.com/shahrzadkhayatbashi/olap-operations4ocel/tree/main/running-example/data}
\item Materials availability: The code and documentation for reproducing the test result are available at \url{https://github.com/shahrzadkhayatbashi/olap-operations4ocel}
\item Code availability: The code is available at \url{https://github.com/jalaliamin/processmining}
\item Author contribution: S.K. conceived the research idea, formalized and implemented the OLAP operations, conducted the case study, and prepared the initial manuscript draft. 
N.M. contributed by developing the data extraction pipelines for the learning management system and assisted in revising and editing the manuscript. 
A.J. supervised the research, validated the implementations and evaluation, and provided critical feedback and revisions to the manuscript. 
All authors read and approved the final version of the paper.
\end{itemize}

\bibliography{sn-bibliography}%

\end{document}